\documentclass[journal=jpccck,manuscript=article]{achemso}
\PassOptionsToPackage{numbers,sort&compress}{natbib}

\usepackage{float}
\usepackage{graphicx}
\usepackage{subcaption}
\usepackage{geometry}
\usepackage{pifont}
\usepackage{amsmath,amsfonts,amssymb}
\usepackage{mathtools}
\usepackage{siunitx}
\usepackage{upgreek}
\usepackage{chemist}
\usepackage{xcolor}
\usepackage[utf8]{inputenc}
\usepackage{textgreek}
\usepackage[normalem]{ulem}
\newcommand{\rev}[1]{#1}
\usepackage{booktabs}
\title{Thermodynamic study of N$_2$ clathrate hydrates from DFT calculations}

\author{L.~Martin-Gondre}
\email{ludovic.martin@univ-fcomte.fr}
\affiliation{Université Marie et Louis Pasteur, CNRS, Institut UTINAM (UMR 6213), équipe ATMOS, F-25000 Besançon, France}

\author{V.~Meko Fotso}
\affiliation{Université Marie et Louis Pasteur, CNRS, Institut UTINAM (UMR 6213), équipe ATMOS, F-25000 Besançon, France}

\author{C.~Métais}
\affiliation{Université Marie et Louis Pasteur, CNRS, Institut UTINAM (UMR 6213), équipe ATMOS, F-25000 Besançon, France}
\alsoaffiliation{Université de Bordeaux, CNRS, Institut des Sciences Moléculaires (UMR 5255), équipe GSM, F-33405 Talence, France}
\alsoaffiliation{Institut Laue Langevin, F-38000 Grenoble, France}

\author{A.~Patt}
\affiliation{Donostia International Physics Center, Universidad del País Vasco / Euskal Herriko Unibertsitatea, E-20018 Donostia-San Sebastián, Spain}

\author{J.~Ollivier}
\affiliation{Institut Laue Langevin, F-38000 Grenoble, France}

\author{A.~Desmedt}
\affiliation{Université de Bordeaux, CNRS, Institut des Sciences Moléculaires (UMR 5255), équipe GSM, F-33405 Talence, France}
\alsoaffiliation{Laboratoire Léon Brillouin, UMR12 CEA-CNRS, F-91190 Gif-sur-Yvette, France}


\begin{document}

\begin{abstract}
Thermodynamic stability of \chemform{N_2} clathrate hydrates in the sI and sII structures is investigated using density functional theory with several exchange–correlation functionals, explicitly accounting for composition (cage occupancies) and pressure at $T=0$~K. Among the tested functionals, revPBE-D3(0) best reproduces experimental lattice parameters and bulk moduli $B_0$. Energetic analyses confirm the strong impact of large-cage double occupancy on sI, whereas the convex-hull results show that sI with single occupancy remains thermodynamically stable up to $\sim\!0.8$~GPa alongside sII with single occupancy. Increasing pressure then stabilizes sII with double occupancy, consistent with its larger large-cage volume and lower framework strain. These results provide a coherent, first-principles thermodynamic framework for \chemform{N_2} hydrate stability and a baseline for finite-temperature extension.
\end{abstract}

\section{Introduction}
\label{Introduction}

Clathrate hydrates are crystalline inclusion compounds in which guest molecules, typically gases, are encapsulated within hydrogen-bonded water cages~\cite{Sloan2008,Broseta2017}. As non-stoichiometric compounds, gas-hydrate cages can accommodate single, multiple, or even no guest molecules, provided that a sufficient number are trapped to stabilize the host framework.

Among the different gas hydrates, nitrogen plays a particularly intriguing role due to its ability to crystallize into two distinct cubic structures: type I (sI) and type II (sII)~\cite{Sloan2008,Petuya2018a}. The sI clathrate unit cell consists of 46 water molecules forming two small cages (SCs) and six large cages (LCs), with a lattice parameter of approximately \SI{12}{\text{\AA}}. In contrast, the sII clathrate unit cell contains 136 water molecules forming sixteen small and eight large cages, resulting in a larger lattice parameter of about \SI{17}{\text{\AA}}~\cite{Sloan2008}.

At thermodynamic equilibrium, nitrogen hydrate is stable in the sII structure at pressures below \SI{<1000}{bar} and near-ambient temperatures, while the sI structure becomes more stable under higher pressures~\cite{Kuhs1997,Sugahara2002,Lundgaard1992}. However, experimental studies show that sI nitrogen hydrate typically forms first at moderate pressures and subsequently transforms into the thermodynamically favored sII phase after a few days~\cite{Petuya2018a}. A similar kinetic effect has been reported for CO clathrate hydrates~\cite{Zhu2014,Petuya2017}, where the sI structure is preferentially formed during the initial stages, whereas the sII structure represents the thermodynamic ground state under equilibrium conditions.

Both SCs and LCs can be occupied by N$_2$ molecules. Notably, experimental studies have demonstrated that two N$_2$ molecules can simultaneously occupy the LCs under elevated pressures in both sI and sII structures~\cite{Chazallon2002,Sasaki2003,Petuya2018a,Kuhs1997,Petuya2018b,Qin2014,Hansen2016}. These findings are consistent with molecular dynamics simulations~\cite{Klaveren2001a,Klaveren2001b} and Grand Canonical Monte Carlo (GCMC) calculations~\cite{Patt2018,Ballenegger2019}. Density functional theory (DFT) studies have also revealed that LC double occupancy plays a critical role in explaining the enhanced thermodynamic stability of the sII structure in CO and N$_2$ hydrates compared to the metastable sI phase~\cite{Zhu2014,Petuya2019,Metais2021}.

First-principles calculations, particularly DFT-based simulations, have emerged as powerful tools for investigating intermolecular interactions, structural properties, and thermodynamic properties in large systems such as gas hydrates. Over the last decade, numerous studies have applied DFT to explore clathrate hydrates with CH$_4$, CO$_2$, N$_2$, CO, H$_2$ and other guest species, providing valuable insights into their structural properties and stability domains~\cite{Desmedt2015,Petuya2019,Metais2021,Cox2014,Sun2017,CabreraRamirez2021a,CabreraRamirez2021b,Omran2022,Sofla2023}. Two types of intermolecular interactions are essential for accurately describing these systems: (i) hydrogen bonding between water molecules forming the host lattice (host–host interactions) and (ii) dispersion forces between guest and water molecules (guest–host interactions). Correctly accounting for van der Waals (vdW) dispersion interactions in condensed phases remains a central challenge in DFT~\cite{Burke2012,Klimes2012,Gillan2016}. Among the available approaches, DFT-D corrections and nonlocal vdW-DF functionals have been widely used to study clathrate hydrates~\cite{Cox2014,CabreraRamirez2021a,CabreraRamirez2021b,Petuya2019,Metais2021,Omran2022}. In earlier studies, DFT calculations for N$_2$ hydrates were performed by testing two functionals: the well-known PBE~\cite{Perdew1996} functional in which vdW interactions are not included, and the original nonlocal vdW-DF development of Dion \emph{et al.}~\cite{Dion2004}. These results revealed a strong dependence of the lattice parameter on cage occupancy, particularly in sI, and highlighted the stabilizing effect of LC double occupancy in sII. However, to achieve a more realistic description of experimental conditions, it is essential to explicitly consider thermodynamic effects, including composition, pressure, and temperature, which are key variables governing clathrate stability.

The present work aims to investigate the thermodynamic properties of N$_2$ clathrate hydrates by explicitly accounting for the effects of composition and pressure. Building on recent studies by Prosmiti and co-workers, which systematically evaluated a wide range of exchange–correlation functionals for CO$_2$ hydrates across sI, sII, and sH structures~\cite{CabreraRamirez2021a,CabreraRamirez2021b,CabreraRamirez2022}, we pursue three main objectives:

\begin{itemize}
\item Evaluate the performance of several DFT functionals in predicting the structural properties of N$_2$ hydrates across different compositions and pressures, benchmarking against experimental data.
\item Analyze guest–host and host–host interaction energies across the various functionals and assess the influence of pressure on their evolution.
\item Establish the thermodynamic stability domains of sI and sII N$_2$ hydrates by constructing convex hull diagrams (enthalpy vs. composition) at multiple pressures, thereby deriving the \SI{0}{\kelvin} phase diagram.
\end{itemize}

The third objective represents a critical step toward establishing a first-principles thermodynamic framework for gas-hydrate stability.

\section{Computational details}
\label{Computdetails}

\textbf{DFT setup.} Electronic structure calculations based on density functional theory (DFT) were performed using the \emph{Vienna Ab initio Simulation Package} (VASP 5.4), a periodic plane-wave basis set code~\cite{Kresse1993,Kresse1996a,Kresse1996b,Kresse1999}, employing the \emph{projector augmented-wave} (PAW) method~\cite{Bloechl1994,Kresse1999} to describe the interaction between valence and core electrons. Different exchange–correlation functionals within the generalized gradient approximation (GGA) were tested:
(i) the semi-local PBE functional~\cite{Perdew1996};
(ii) the non-local vdW-DF functional~\cite{Dion2004};
(iii) the improved vdW-DF2 functional~\cite{Lee2010}; and
(iv) the revPBE-D3(0) functional, which includes an empirical DFT-D dispersion correction and has been shown to provide accurate results for describing CO$_2$--H$_2$O interactions in gas hydrates~\cite{CabreraRamirez2021a}. The plane-wave basis set was expanded up to an energy cutoff of \SI{520}{\electronvolt} for PBE and vdW-DF calculations and increased to \SI{600}{\electronvolt} for vdW-DF2 and revPBE-D3(0), ensuring accurate convergence of total energies. \rev{Because of the large unit cells of the sI ($a \approx \SI{12}{\text{\AA}}$) and sII ($a \approx \SI{17}{\text{\AA}}$) clathrate hydrates, the simulations were performed using the conventional crystallographic unit cell. Convergence tests using a larger supercell ($2\times2\times2$) resulted in total energy differences below \SI{1}{\milli\electronvolt} per molecule. Likewise, the Brillouin zone was sampled using a Γ-point mesh ($1\times1\times1$), which we verified to be sufficient for energy convergence within 1 meV per molecule compared with denser $2\times2\times2$ and $4\times4\times4$ meshes. These tests confirm that $\Upgamma$-point sampling and the conventional unit cell are sufficient for the structural and thermodynamic properties investigated here.}

\begin{figure}[ht]
  \begin{center}
    \includegraphics[width=0.8\textwidth]{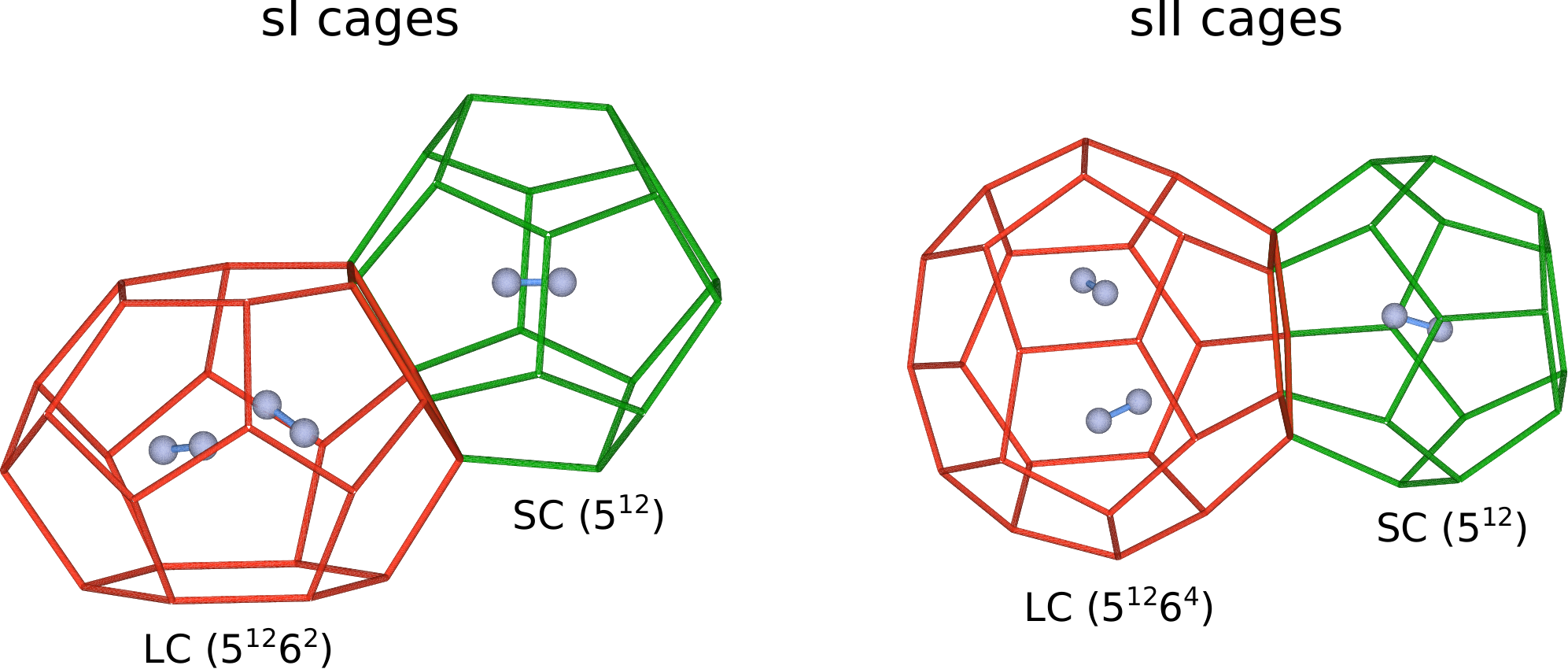}
    \caption{Polyhedral representation of the small ($5^{12}$) and large cages ($5^{12}6^{2}$ for sI and $5^{12}6^{4}$ for sII) in clathrate hydrate structures after structural relaxation. The left panel shows the sI cages, while the right panel shows the sII cages. Small cages (SC) are singly occupied by N$_2$, whereas large cages (LC) are doubly occupied.}
    \label{fig:cages}
  \end{center}
\end{figure}

\medskip
\textbf{Structural relaxation.} Previous calculations reported in~\cite{Metais2021} with PBE and vdW-DF functionals employed a custom program to define hydrogen positions according to the ice rules. In the present work, the initial positions of the water molecules were taken from Takeuchi \emph{et al.}~\cite{Takeuchi2013}, where hydrogen positions were optimized to yield zero dipole moments while satisfying the ice rules within the unit cell. \rev{The guest nitrogen molecules were initially positioned at the center of the cages} (see Figure~\ref{fig:cages}). The resulting structures are consistent between both approaches, particularly in terms of interaction energies, although small differences in structural parameters may arise, as discussed in the Results section.

Structural optimizations were performed by relaxing all atomic positions until forces on each atom were below \SI{0.01}{\electronvolt\per\text{\AA}}. To determine the equilibrium lattice parameter $a_0$ and bulk modulus $B_0$, structural optimizations were carried out for approximately 15 different lattice constants, varying by \SI{0.10}{\text{\AA}} for sI and \SI{0.15}{\text{\AA}} for sII, for each exchange–correlation functional. The DFT energies were then fitted to the isothermal third-order Birch--Murnaghan equation of state (BM EOS)~\cite{Birch1947}:

\begin{equation}
  E(V) = E_0 + \frac{9V_0B_0}{16} \left\{\left[\left(\frac{V_0}{V}\right)^{2/3} - 1\right]^3 B'_0
    + \left[\left(\frac{V_0}{V}\right)^{2/3} - 1\right]^2\left[6 - 4\left(\frac{V_0}{V}\right)^{2/3}\right]\right\}
  \label{eq:EV}
\end{equation}

where $E_0$ is the minimum energy associated with the equilibrium volume $V_0$ ($V_0=a_0^3$), $B_0$ is the bulk modulus at \SI{0}{\kelvin} and \SI{0}{\giga\pascal}, and $B'_0$ is its pressure derivative at \SI{0}{\kelvin}. From the fitted parameters in Eq.~\ref{eq:EV}, the corresponding pressure--volume relationship is obtained as:

\begin{equation}
  P(V) = \frac{3B_0}{2}\left[\left(\frac{V_0}{V}\right)^{7/3} - \left(\frac{V_0}{V}\right)^{5/3}\right]
  \left\{ 1 + \frac{3}{4}\left(B_0' - 4\right)\!\left[ \left(\frac{V_0}{V}\right)^{2/3} - 1 \right]\right\}
  \label{eq:PV}
\end{equation}

Volume relaxations using the BM EOS were performed for different cage occupancies in order to investigate the influence of composition on nitrogen hydrate properties. $\theta_{\mathrm{SC}}$ and $\theta_{\mathrm{LC}}$ represent the occupancies of the small and large cages, respectively. Three types of filling were considered:
(i) single occupancy with $\theta_{\mathrm{SC}}=\theta_{\mathrm{LC}}=1$;
(ii) double occupancy of the large cages with $\theta_{\mathrm{SC}}=1$ and $1<\theta_{\mathrm{LC}}\leq2$; and
(iii) under-occupancy with $\theta_{\mathrm{SC}}<1$ for $\theta_{\mathrm{LC}}=1$ and $\theta_{\mathrm{LC}}<1$ for $\theta_{\mathrm{SC}}=1$. \rev{For double occupancy, the \chemform{N_2} molecules were initially positioned symmetrically in the large cages, such that the center of mass of the two molecules coincides with the center of the cage, with an angle between the molecular axes of approximately \SI{45}{\degree} (Figure~\ref{fig:cages}).}

\medskip
\textbf{Intermolecular energies.} The energetics of the sI and sII phases were analyzed by determining the guest--host interaction energy $E^{GH}$, which corresponds to the binding energy per N$_2$ molecule encapsulated in the water framework, and the host--host energy $E^{HH}$, which represents the cohesive energy per H$_2$O molecule of the empty clathrate:

\begin{equation}
 E^{GH}_{\mathrm{sI}} = \frac{E_{\mathrm{sI}} - E_{\mathrm{sI}}^{\mathrm{empty}} - (2\theta_{\mathrm{SC}} + 6\theta_{\mathrm{LC}})E_{\mathrm{N_2}}}{(2\theta_{\mathrm{SC}} + 6\theta_{\mathrm{LC}})}
 \label{eq:EGHsI}
\end{equation}

\begin{equation}
 E^{GH}_{\mathrm{sII}} = \frac{E_{\mathrm{sII}} - E_{\mathrm{sII}}^{\mathrm{empty}} - (16\theta_{\mathrm{SC}} + 8\theta_{\mathrm{LC}})E_{\mathrm{N_2}}}{(16\theta_{\mathrm{SC}} + 8\theta_{\mathrm{LC}})}
 \label{eq:EGHsII}
\end{equation}

\begin{equation}
 E^{HH}_{\mathrm{sI}} = \frac{E_{\mathrm{sI}}^{\mathrm{empty}} - 46E_{\mathrm{H_2O}}^{\mathrm{(ice)}}}{46}
 \label{eq:EHHsI}
\end{equation}

\begin{equation}
 E^{HH}_{\mathrm{sII}} = \frac{E_{\mathrm{sII}}^{\mathrm{empty}} - 136E_{\mathrm{H_2O}}^{\mathrm{(ice)}}}{136}
 \label{eq:HHsII}
\end{equation}

where $E_{\mathrm{sI}}$ and $E_{\mathrm{sII}}$ denote the DFT energies of the sI and sII N$_2$ hydrates, respectively. $E_{\mathrm{sI}}^{\mathrm{empty}}$ and $E_{\mathrm{sII}}^{\mathrm{empty}}$ are the total energies of the corresponding empty hydrate frameworks, obtained by removing the N$_2$ molecules of the water lattice without further relaxation. $E_{\mathrm{N_2}}$ is the total energy of an isolated N$_2$ molecule in its ground state, while $E_{\mathrm{H_2O}}^{\mathrm{(ice)}}$ is the energy per H$_2$O molecule in the relaxed hexagonal ice I$_\mathrm{h}$ crystal.

Calculations for an isolated N$_2$ molecule were performed in a cubic simulation cell with an edge length of approximately \SI{10}{\text{\AA}}. For hexagonal ice, calculations were carried out using a 4-molecule unit cell in the proton-ordered phase, sampled with a $6\times6\times6$ $\Gamma$-centered k-point mesh. Volume relaxation of the ice crystal was performed by isotropically varying the lattice parameters, resulting in $c/a$ ratio of approximately $1.63$, consistent with experimental values~\cite{Santra2013}.

\rev{The guest--guest interaction energy was also evaluated using a frozen-host energy decomposition scheme:}

\rev{\begin{equation}
 E^{GG} = \frac{E_{\mathrm{clath}} - E_{\mathrm{clath}}^{A} - E_{\mathrm{clath}}^{B} + E_{\mathrm{clath}}^{0}}{N_{\mathrm{pair}}}
 \label{eq:EGG}
\end{equation}}

\rev{where $E_{\mathrm{clath}}$ is the total energy of the sI or sII hydrate structure, $E_{\mathrm{clath}}^{A}$ and $E_{\mathrm{clath}}^{B}$ are the total energies of the two complementary partially emptied structures. $E_{\mathrm{clath}}^{0}$ is the total energy of the reference structure in which both guest molecules forming each considered pair are removed from the hydrate, while all other guest molecules remain unchanged, without further relaxation. $N_{\mathrm{pair}}$ denotes the number of guest pairs included in the decomposition.}

\rev{For doubly occupied large cages, the pair corresponds to the two \chemform{N_2} molecules confined in the same cage, so that $E^{GG}$ measures an effective intra-cage guest--guest interaction. For singly
occupied structures, the same expression can be applied to a selected pair of guests located in different cages, allowing the evaluation of inter-cage guest--guest interactions, which are expected to be significantly weaker.}

\medskip
\textbf{Convex hull.}
The stability of the sI and sII N$_2$ clathrate hydrates was evaluated from their non-bonding (cohesive) energy, $E^{NB}$, which accounts for all intermolecular interactions within the system. It was calculated as the energy difference between the hydrate and its separate components, normalized per molecule:

\begin{equation}
 E^{NB}_{\mathrm{sI}} = \frac{E_{\mathrm{sI}} - (2\theta_{\mathrm{SC}} + 6\theta_{\mathrm{LC}})E_{\mathrm{N_2}} - 46E_{\mathrm{H_2O}}^{\mathrm{(ice)}}}{2\theta_{\mathrm{SC}} + 6\theta_{\mathrm{LC}} + 46}
 \label{eq:NBsI}
\end{equation}

\begin{equation}
 E^{NB}_{\mathrm{sII}} = \frac{E_{\mathrm{sII}} - (16\theta_{\mathrm{SC}} + 8\theta_{\mathrm{LC}})E_{\mathrm{N_2}} - 136E_{\mathrm{H_2O}}^{\mathrm{(ice)}}}{16\theta_{\mathrm{SC}} + 8\theta_{\mathrm{LC}} + 136}
 \label{eq:NBsII}
\end{equation}

Taking into account the effect of pressure $P$ requires computing the enthalpy, defined as $H=E+PV$. Thermodynamic stability was then assessed by computing the relative enthalpy per molecule at multiple pressures and cage occupancies, as follows:

\begin{equation}
 \Delta H_{\mathrm{sI,sII}} = 
 \frac{H_{\mathrm{sI,sII}} - n_{\mathrm{N_2}}H_{\mathrm{N_2}} - n_{\mathrm{H_2O}}H_{\mathrm{H_2O}}^{\mathrm{(ice)}}}
 {n_{\mathrm{N_2}} + n_{\mathrm{H_2O}}},
 \label{eq:DeltaH}
\end{equation}

where $(n_{\mathrm{N_2}}, n_{\mathrm{H_2O}}) = (2\theta_{\mathrm{SC}} + 6\theta_{\mathrm{LC}}, 46)$ for sI and 
$(16\theta_{\mathrm{SC}} + 8\theta_{\mathrm{LC}}, 136)$ for sII. This expression can also be written in compact form as:

\begin{equation}
 \Delta H = E^{NB} + P\Delta V,
 \label{eq:DeltaHcompact}
\end{equation}

where $\Delta V$ denotes the volume difference per molecule between the hydrate and the separated constituents at pressure $P$.

To properly account for the thermodynamic conditions, i.e. at $T=\SI{0}{\kelvin}$, the enthalpies of solid N$_2$ and ice were computed using their thermodynamically stable phases at each pressure. For N$_2$, the $\alpha$, $\gamma$ and $\epsilon$ phases were included, as they correspond to the stable molecular solids at low temperature for pressures between $0$ and $\SI{18}{\giga\pascal}$~\cite{Tassini2005}. For H$_2$O, the hydrogen-ordered ices XI, II, XV, and VIII were considered, with ice VIII remaining stable up to about \SI{50}{\giga\pascal}~\cite{Salzmann2019}. The computational setup for these molecular solids was identical to that used for ice I$_\mathrm{h}$.

Finally, convex hull diagrams (enthalpy vs. composition) were constructed using the relative enthalpies $\Delta H$. This analysis enabled the identification of the thermodynamically stable phases at each pressure and allowed the prediction of a compositional phase diagram of N$_2$ clathrate hydrates at $\SI{0}{\kelvin}$. The results highlight the key role of large-cage occupancy in stabilizing clathrate structures.

\section{Results and discussion}
\label{Results and discussion}

\subsection{Lattice parameters and bulk modulus}
\begin{figure}[ht]
  \begin{center}
    \includegraphics[width=0.8\textwidth]{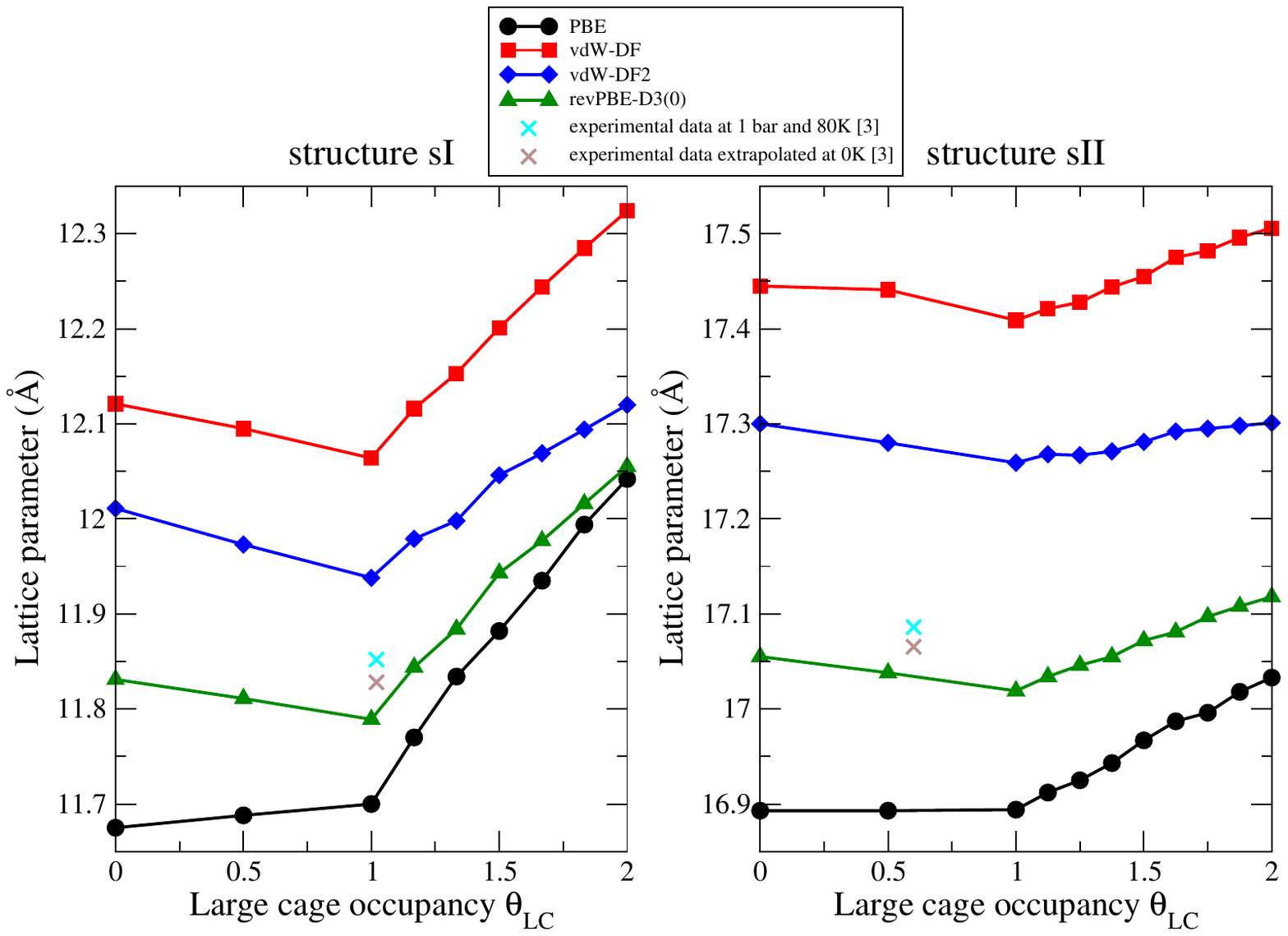}
    \caption{Lattice parameter as a function of large cage occupancy for sI (left) and sII (right) structures, obtained with various DFT functionals: PBE (black circles), vdW-DF (red squares), vdW-DF2 (blue diamonds), and revPBE-D3(0) (green triangles). Crosses indicate experimental values measured at $80$ K and at $0$ K (the latter extrapolated from the $80$ K data according to Ref.~\cite{Petuya2018a}), both at atmospheric pressure (color-coded as indicated in the figure).}
    \label{fig:unitcell}
  \end{center}
\end{figure}

The fitting of DFT data to the BM EOS for different large-cage occupancies ($\theta_{\mathrm{SC}}=1$) is shown in Figure~\ref{fig:unitcell}, where the equilibrium lattice parameter $a_0$ is plotted against the large cage occupancy $\theta_{\mathrm{LC}}$ for both sI and sII structures. The experimental cell parameter obtained from neutron diffraction is also reported at atmospheric pressure and $\SI{80}{\kelvin}$~\cite{Petuya2018a}, together with its extrapolated value at $\SI{0}{\kelvin}$ using the temperature dependence of the lattice constant in ref.~\cite{Petuya2018a}, for better comparison with DFT results. The revPBE-D3(0) functional yields excellent agreement with experimental data for both structures, with relative deviations below 0.3\%. In contrast, the vdW-DF functional significantly overestimates the lattice parameter due to its overly repulsive character, as discussed previously~\cite{Metais2021,Cox2014}, while the vdW-DF2 version improves the structural description but still overshoots the experimental values. Conversely, PBE systematically underestimates the cell parameter, mainly due to its tendency to overbind hydrogen bonds in the water framework, leading to shorter O···O distances~\cite{Cox2014,Gillan2016}.

The evolution of the lattice parameter with $\theta_{\mathrm{LC}}$ for the different functionals reveals two main regimes: (i) a slight decrease in the under-occupancy domain ($\theta_{\mathrm{LC}}\leqslant 1$), and (ii) a pronounced increase for $\theta_{\mathrm{LC}} > 1$, particularly in the sI structure and to a lesser extent in sII. An increase of guest content in the under-occupied region enhances guest-host interactions, leading to a slight cage contraction. In contrast, double occupancy produces the opposite effect: steric repulsion expands the lattice, the effect being less pronounced in sII owing to its larger large cages. An exception to this general behavior is observed with PBE, where the lack of dispersion corrections results in a monotonic lattice expansion with occupancy, accompanied by larger variations.

Compared to previous work~\cite{Metais2021}, the optimized structures obtained here from initial atomic positions of Takeuchi \emph{et al.}~\cite{Takeuchi2013} exhibit lower lattice parameters and a more monotonic evolution with cage occupancy. This confirms a more reliable optimization of the hydrogen-bonded water substructure, leading to a smoother variation of the cell parameter with increasing N$_2$ content, even though the global occupancy trends are preserved.

\begin{figure}[ht]
  \begin{center}
    \includegraphics[width=0.8\textwidth]{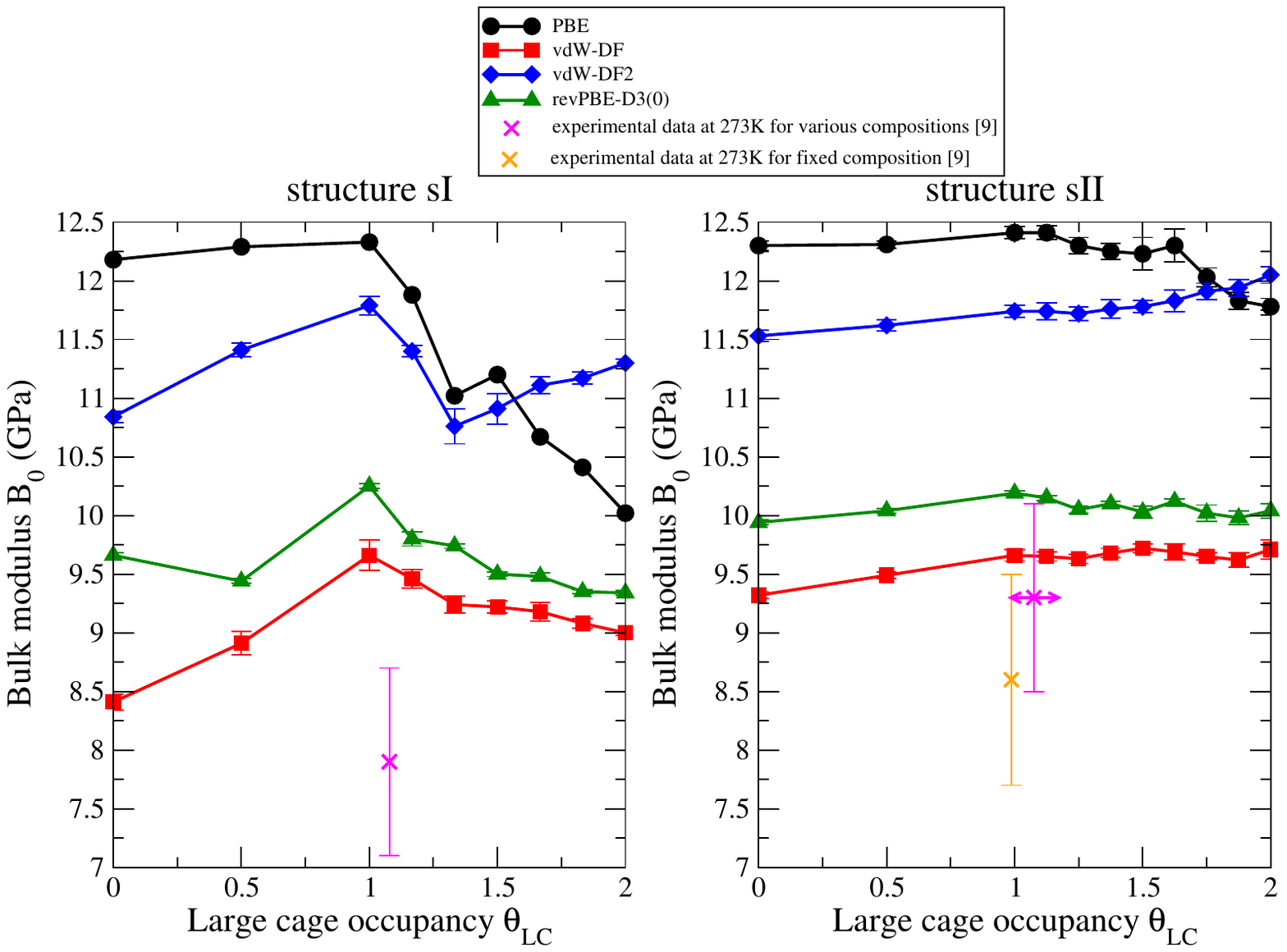}
    \caption{Bulk modulus $B_0$ as a function of the large cage occupancy for sI (left) and sII (right) structures obtained with various DFT functionals: PBE (black circles), vdW-DF (red squares), vdW-DF2 (blue diamonds), and revPBE-D3(0) (green triangles). Crosses indicate experimental values measured at $273$ K for several compositions (purple crosses) and for a fixed composition (orange crosses)~\cite{Chazallon2002}, see the text for details. Error bars are also indicated for DFT and experimental values.}
    \label{fig:bulkmodulus}
  \end{center}
\end{figure}

The evolution of the bulk modulus $B_0$ with the large-cage occupancy $\theta_{\mathrm{LC}}$ ($\theta_{\mathrm{SC}}=1$) is presented in Figure~\ref{fig:bulkmodulus} for both sI and sII structures. In both cases, $B_0$ increases with increasing occupancy in the under-occupancy region ($\theta_{\mathrm{LC}}<1$), reflecting the progressive stiffening of the host lattice, in agreement with the structural compression observed in Figure~\ref{fig:unitcell}. Conversely, in the double-occupancy domain ($\theta_{\mathrm{LC}}>1$), $B_0$ decreases almost monotonically, following the lattice expansion and indicating a softening of the host framework at higher guest concentrations. This global trend is observed for both structures, although the variations are smoother in sII, consistent with the softer evolution of its lattice parameter (Figure~\ref{fig:unitcell}) resulting from its larger cage size and lower structural sensitivity to guest filling.

Among the exchange–correlation functionals, revPBE-D3(0) and vdW-DF provide the best agreement with experimental measurements at \SI{273}{\kelvin}~\cite{Chazallon2002}, yielding bulk moduli close to 9–10~GPa for the singly occupied structures, whereas PBE and vdW-DF2 systematically overestimate $B_0$. Experimental data were obtained for two distinct conditions in the sII phase: (i) various compositions in the range $0.98\leqslant\theta_{\mathrm{LC}}\leqslant1.17$ ($0.82\leqslant\theta_{\mathrm{SC}}\leqslant0.96$), and (ii) a fixed composition $\theta_{\mathrm{LC}}\approx 0.98$ ($\theta_{\mathrm{SC}}\approx 0.82$), obtained by rapid compression to ensure isosteric conditions~\cite{Chazallon2002}. For the sI structure, the DFT-calculated bulk moduli are higher but remain close to the experimental values obtained for sII, particularly when the uncertainty ranges are considered. As underlined by Vlasic \emph{et al.}~\cite{Vlasic2016}, lower experimental bulk moduli compared to theoretical values are expected. Indeed, DFT-derived $B_0$ values correspond to an ideal single crystal without defects, whereas experimental samples are polycrystalline hydrates that naturally contain imperfections. In addition, the calculations are performed at \SI{0}{\kelvin}, while experiments are carried out at 273~K. Since increasing temperature generally leads to a decrease in elastic stiffness, the DFT results can be regarded as upper-limit values for defect-free monocrystals. Despite these differences, revPBE-D3(0) and vdW-DF provide satisfactory agreement with the experimental data, particularly for the sII structure.

\begin{figure}[ht]
  \centering
    \begin{subfigure}[b]{0.49\textwidth}
        \centering
        \includegraphics[trim=20 30 80 60, clip, width=\textwidth]{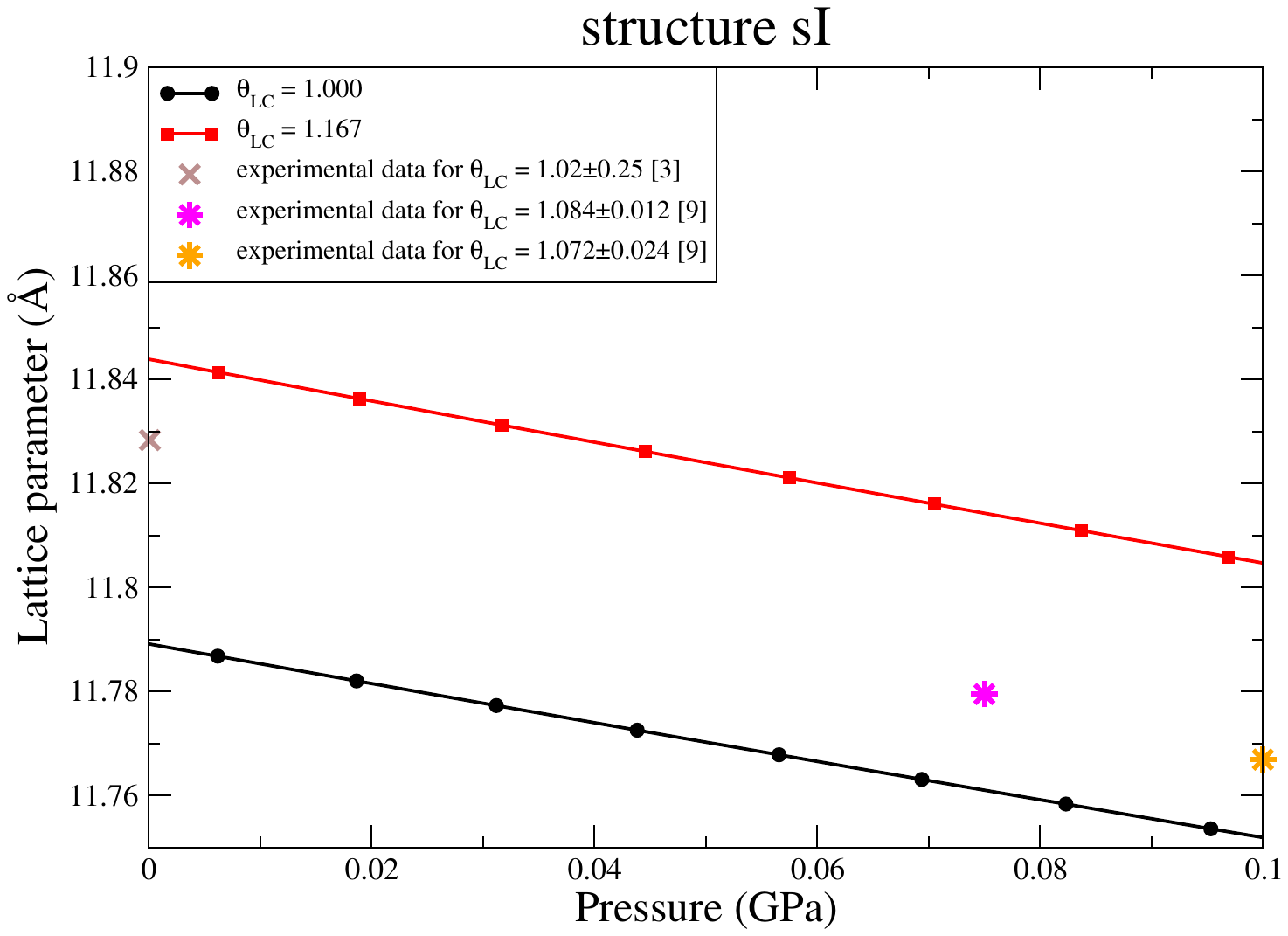}
        \label{fig:unitcellPresI}
    \end{subfigure}
    \hfill
    \begin{subfigure}[b]{0.49\textwidth}
        \centering
        \includegraphics[trim=20 30 80 60, clip, width=\textwidth]{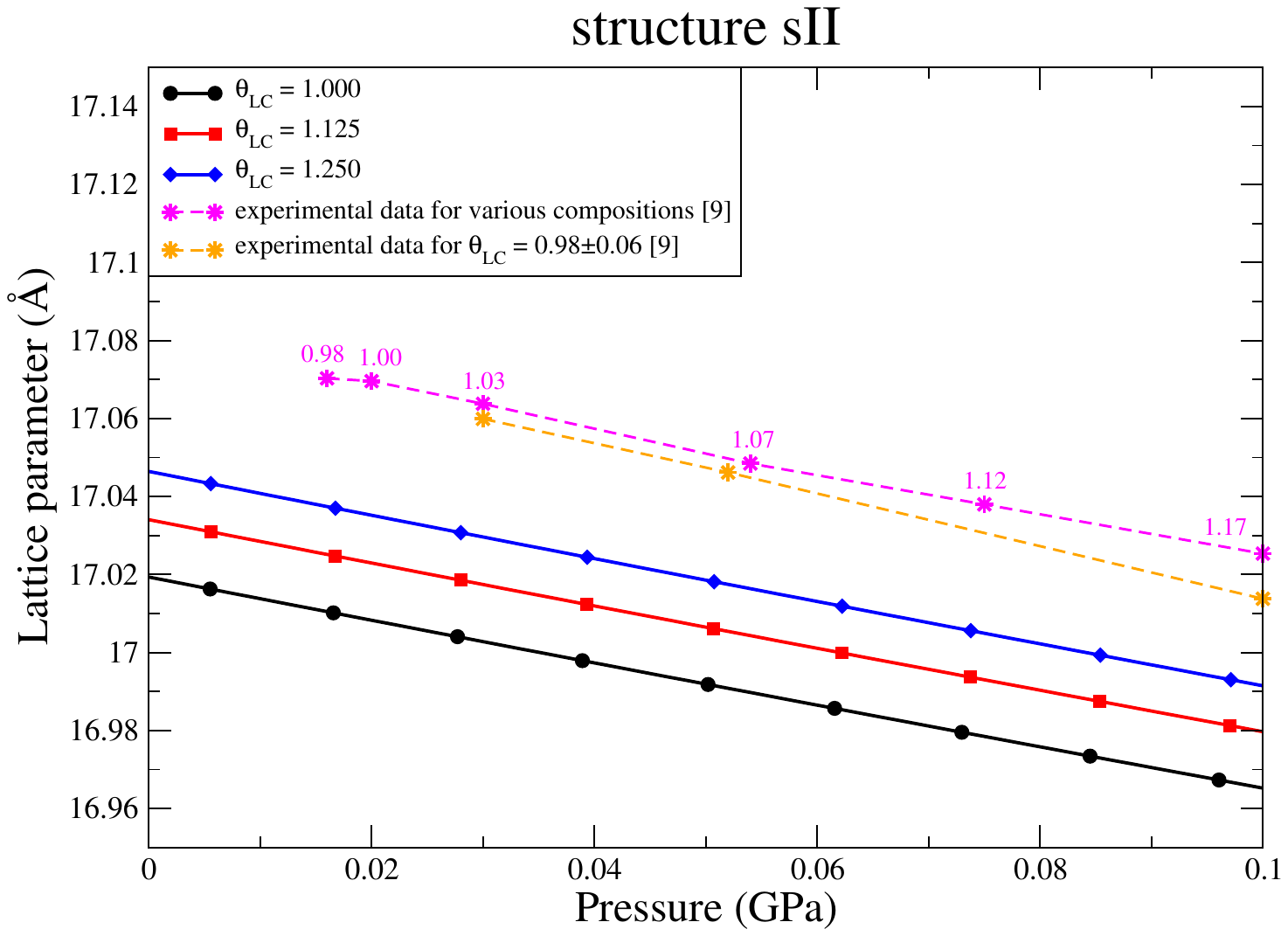}
        \label{fig:unitcellPresII}
    \end{subfigure}
    \caption{Lattice parameter as a function of pressure in GPa for sI (left) and sII (right) structures, obtained with revPBE-D3(0) functional for different large cage occupancies $\theta_{LC}$. Symbols indicate experimental values extrapolated to \SI{0}{\kelvin} according to Ref.~\cite{Petuya2018a}. Large cage occupancies are indicated in the caption and directly on the figure for the sII various composition case (purple stars).}
    \label{fig:unitcellPres}
\end{figure}

The effect of pressure on the lattice parameter can be studied using Eq.~\ref{eq:PV}, which then allows one to evaluate not only the bulk modulus $B_0$ but also its first pressure derivative, $B_0'$. Figure~\ref{fig:unitcellPres} reports the evolution of the lattice parameter with pressure for both structures obtained with revPBE-D3(0), together with a comparison to experimental data~\cite{Chazallon2002,Petuya2018a}. Among the tested functionals, revPBE-D3(0) has been retained as the most accurate choice, combining an excellent description of the equilibrium lattice parameter (Figure~\ref{fig:unitcell}) with bulk moduli $B_0$ in close agreement with experiments (Figure~\ref{fig:bulkmodulus}). Note that fitted $B_0'$ values for revPBE-D3(0) lie between 5 and 6 (comparable to those obtained with PBE), whereas the vdW-DF family gives slightly smaller values, between 4.5 and 5.5 (all BM EOS fitted parameters are provided in the Supporting Information for all functionals).

As in Figure~\ref{fig:unitcell}, the lattice parameters computed with revPBE-D3(0) remain systematically lower than those obtained with the other functionals, consistent with its slightly stiffer framework. For the sI structure, DFT results show very good agreement with experiments, since the two occupancies considered here (single occupancy, $\theta_{\mathrm{LC}}=1$, and one doubly occupied large cage, $\theta_{\mathrm{LC}}=1.167$) frame the experimental data measured for $1.02\leqslant\theta_{\mathrm{LC}}\leqslant1.084$~\cite{Chazallon2002,Petuya2018a}. In addition, the pressure dependence of $a$ exhibits an almost identical slope to that observed experimentally, confirming that the compressibility and $B_0'$ values are realistically reproduced.

For the sII structure, the DFT-derived lattice parameters do not perfectly reproduce the absolute experimental values even when considering higher occupancies up to $\theta_{\mathrm{LC}}=1.25$ (two doubly occupied LC), although the relative deviation never exceeds 0.3\%. The evolution of $a$ with pressure follows the experimental trend with similar slopes, except for the fixed-composition measurements (orange crosses) where the experimental curve shows a slightly higher slope, reflecting a lower resistance to compression and hence a smaller $B_0'$. Overall, these results reinforce the consistency between the structural and mechanical descriptions: DFT using revPBE-D3(0) provides an accurate representation of both equilibrium geometry and pressure response for N$_2$ clathrate hydrates.

\rev{Additional microscopic insight is provided by the analysis of the \chemform{N_2} center-of-mass (COM) positions in the small cages (SC) and large cages (LC) as a function of composition and pressure (Table~\ref{table_revPBE}). In the sI structure, singly occupied SC remain essentially centered, with COM deviations below 0.3\% over the whole pressure range. Singly occupied LC are also only weakly off-center, with average deviations of about 1.7--1.9\% for $\theta_{\mathrm{LC}}=1.0$, showing that a single \chemform{N_2} molecule remains close to the cage center. In the sII structure, SC also remain close to the center, although with slightly larger deviations ($\sim 1$\%), whereas singly occupied LC exhibit significantly larger off-center displacements, reaching about 11 $\pm$ 4\% for $\theta_{\mathrm{LC}}=1.5$ at zero pressure. This behavior reflects the larger free volume available in the sII large cages, which allows the guest molecule to explore more eccentric positions.}

\rev{The effect of double occupancy is even more striking. In both sI and sII, the COM deviations of the two \chemform{N_2} molecules in doubly occupied LC are remarkably similar, remaining close to 34--36\% of the cage radius over the whole pressure range. This indicates that, once two molecules occupy the same cage, they adopt nearly symmetric off-center positions in order to reduce steric crowding and guest--guest repulsion, while maintaining the center of mass of the pair close to the cage center. Pressure has only a weak effect on these deviations.}

\rev{A marked difference between sI and sII appears in the orientational arrangement of the two \chemform{N_2} molecules inside doubly occupied LC. In sI, the relative angle $\alpha$ is generally closer to parallel configurations, with $\alpha < 35^\circ$ for $\theta_{\mathrm{LC}}=1.5$ and decreasing further with pressure and occupancy, down to quasi-parallel arrangements for $\theta_{\mathrm{LC}}=2.0$ above about 1~GPa. This trend is accompanied by a gradual reduction of the intermolecular distance $d$, from about 3.08~\AA\ at ambient pressure to 2.96~\AA\ near 2~GPa. The smaller size of the sI large cages therefore imposes stronger geometric constraints, favoring alignment of the molecular axes under compression. In contrast, the larger LC of sII allow much more tilted configurations, with $\alpha$ typically in the 45--70$^\circ$ range and only a limited pressure dependence. The intermolecular distance remains systematically larger than in sI, consistent with the larger free volume available in sII. The relatively large standard deviations of $\alpha$, especially in sII, further indicate a broader distribution of accessible orientations in the less confined cages.}

\rev{Overall, these results show that cage size primarily controls the guest positioning and orientational freedom, with sI promoting more constrained and aligned configurations under pressure, whereas sII allows larger positional and orientational flexibility due to its larger cavity size.}

\newcommand{\NA}{\multicolumn{1}{c}{--}}
\setlength{\tabcolsep}{3pt}
\begin{table*}[t]
  \centering
  \caption{Center-of-mass deviations of \chemform{N_2} molecules in sI and sII clathrate cages obtained with the revPBE-D3(0) functional, expressed as a percentage of the cage radius (0\%: cage center; 100\%: cage wall) for various $\theta_{LC}$ occupancy and pressure. Values are reported as mean $\pm$ standard deviation. SC and LC denote small and large cages, while SO and DO refer to single and double occupancy, respectively. For doubly occupied cages, $\alpha$ is the relative orientation of the two \chemform{N_2} molecules (\SI{0}{\degree}: parallel axes; \SI{90}{\degree}: perpendicular axes), and $d$ their intermolecular distance.}
  \label{table_revPBE}
  \footnotesize
  \begin{tabular}{@{}
      S[table-format=1.1]
      S[table-format=1.2]
      S[table-format=1.2(1)]
      S[table-format=1.1(2)]
      S[table-format=2.2(2)]
      S[table-format=2.1(2)]
      S[table-format=2.2(1)]
      S[table-format=1.2]
      S[table-format=1.2(1)]
      S[table-format=2.2(2)]
      S[table-format=2.2(2)]
      S[table-format=2.1(2)]
      S[table-format=2.2(1)]
    @{}}
    \toprule
    \multicolumn{1}{c}{} &
    \multicolumn{6}{c}{\textbf{structure sI}} &
    \multicolumn{6}{c}{\textbf{structure sII}} \\
    \cmidrule(lr){2-7}
    \cmidrule(lr){8-13}
    \multicolumn{2}{c}{} &
    \multicolumn{3}{c}{\textbf{COM dev. (\%)}} &
    \multicolumn{2}{c}{\textbf{\chemform{N_2} DO}} &
    \multicolumn{1}{c}{} &
    \multicolumn{3}{c}{\textbf{COM dev. (\%)}} &
    \multicolumn{2}{c}{\textbf{\chemform{N_2} DO}}\\
    \cmidrule(lr){3-5}
    \cmidrule(lr){6-7}
    \cmidrule(lr){9-11}
    \cmidrule(lr){12-13}

    {$\theta_{\mathrm{LC}}$} &
    {$\mathrm{P_{sI}}$ (\si{\giga\pascal})} & {SC} & {LC-SO} & {LC-DO} & {$\alpha$ (\si{\degree})} & {$d$ (\si{\AA})} &
    {$\mathrm{P_{sII}}$ (\si{\giga\pascal})} & {SC} & {LC-SO} & {LC-DO} & {$\alpha$ (\si{\degree})} & {$d$ (\si{\AA})}\\
    \midrule
    1.0 & 0.00 & \num{0.22(4)} & \num{1.9(2)} & \NA & \NA & \NA & 0.00 & \num{1.2(4)} & \num{3.4(9)}  & \NA & \NA & \NA \\
        & 0.57 & \num{0.17(1)} & \num{1.9(2)} & \NA & \NA & \NA & 0.56 & \num{1.0(4)} & \num{3.9(10)} & \NA & \NA & \NA \\
        & 0.95 & \num{0.11(1)} & \num{1.8(2)} & \NA & \NA & \NA & 0.94 & \num{1.0(5)} & \num{4.9(11)} & \NA & \NA & \NA \\
        & 1.88 & \num{0.10(1)} & \num{1.7(3)} & \NA & \NA & \NA & 1.90 & \num{0.8(4)} & \num{5.3(11)} & \NA & \NA & \NA \\
    \midrule
    1.5 & 0.00 & \num{0.21(1)} & \num{3.5(13)}  & \num{35.3(2)} & \num{35(8)}  & \num{3.04(1)} &
          0.00 & \num{1.2(5)}  & \num{11.4(38)} & \num{34.3(5)} & \num{67(20)} & \num{3.16(3)} \\
        & 0.69 & \num{0.25(4)} & \num{3.7(14)}  & \num{35.3(3)} & \num{30(7)}  & \num{2.98(1)} &
          0.67 & \num{1.1(5)}  & \num{11.1(37)} & \num{34.3(5)} & \num{66(20)} & \num{3.10(2)} \\
        & 1.05 & \num{0.25(8)} & \num{3.8(13)}  & \num{35.4(6)} & \num{16(18)} & \num{2.96(2)} &
          1.05 & \num{1.1(5)}  & \num{10.9(36)} & \num{34.3(5)} & \num{65(20)} & \num{3.07(2)} \\
        & 1.46 & \num{0.25(6)} & \num{3.8(13)}  & \num{35.5(5)} & \num{16(18)} & \num{2.94(2)} &
          1.50 & \num{1.0(4)}  & \num{10.1(34)} & \num{34.3(5)} & \num{65(20)} & \num{3.05(3)} \\
        & 1.92 & \num{0.29(3)} & \num{4.0(14)}  & \num{35.6(7)} & \num{24(22)} & \num{2.93(3)} &
          2.00 & \num{1.0(3)}  & \num{8.8(32)}  & \num{34.4(4)} & \num{64(21)} & \num{3.03(2)} \\
    \midrule
    2.0 & 0.00 & \num{0.18(3)} & \NA & \num{35.5(4)} & \num{14(15)} & \num{3.08(1)} &
          0.00 & \num{1.1(4)}  & \NA & \num{34.4(4)} & \num{46(25)} & \num{3.18(2)} \\
        & 0.72 & \num{0.17(6)} & \NA & \num{35.6(4)} & \num{9(11)}  & \num{3.02(1)} &
          0.43 & \num{1.2(4)}  & \NA & \num{34.4(4)} & \num{46(24)} & \num{3.14(2)} \\
        & 1.08 & \num{0.22(7)} & \NA & \num{35.7(3)} & \num{4(3)}   & \num{3.00(1)} &
          1.18 & \num{1.2(6)}  & \NA & \num{34.4(4)} & \num{47(23)} & \num{3.08(2)} \\
        & 1.49 & \num{0.11(0)} & \NA & \num{35.7(3)} & \num{3(1)}   & \num{2.98(1)} &
          1.63 & \num{1.1(6)}  & \NA & \num{34.4(4)} & \num{47(23)} & \num{3.06(2)} \\
        & 1.97 & \num{0.07(2)} & \NA & \num{35.7(3)} & \num{3(1)}   & \num{2.96(1)} &
          2.16 & \num{1.2(7)}  & \NA & \num{34.5(4)} & \num{45(22)} & \num{3.04(2)} \\
    \midrule
    \bottomrule
  \end{tabular}
\end{table*}

\subsection{Intermolecular energies}
\begin{figure}[ht]
  \centering
    \begin{subfigure}[b]{0.49\textwidth}
        \centering
        \includegraphics[trim=15 25 75 90, clip, width=\textwidth]{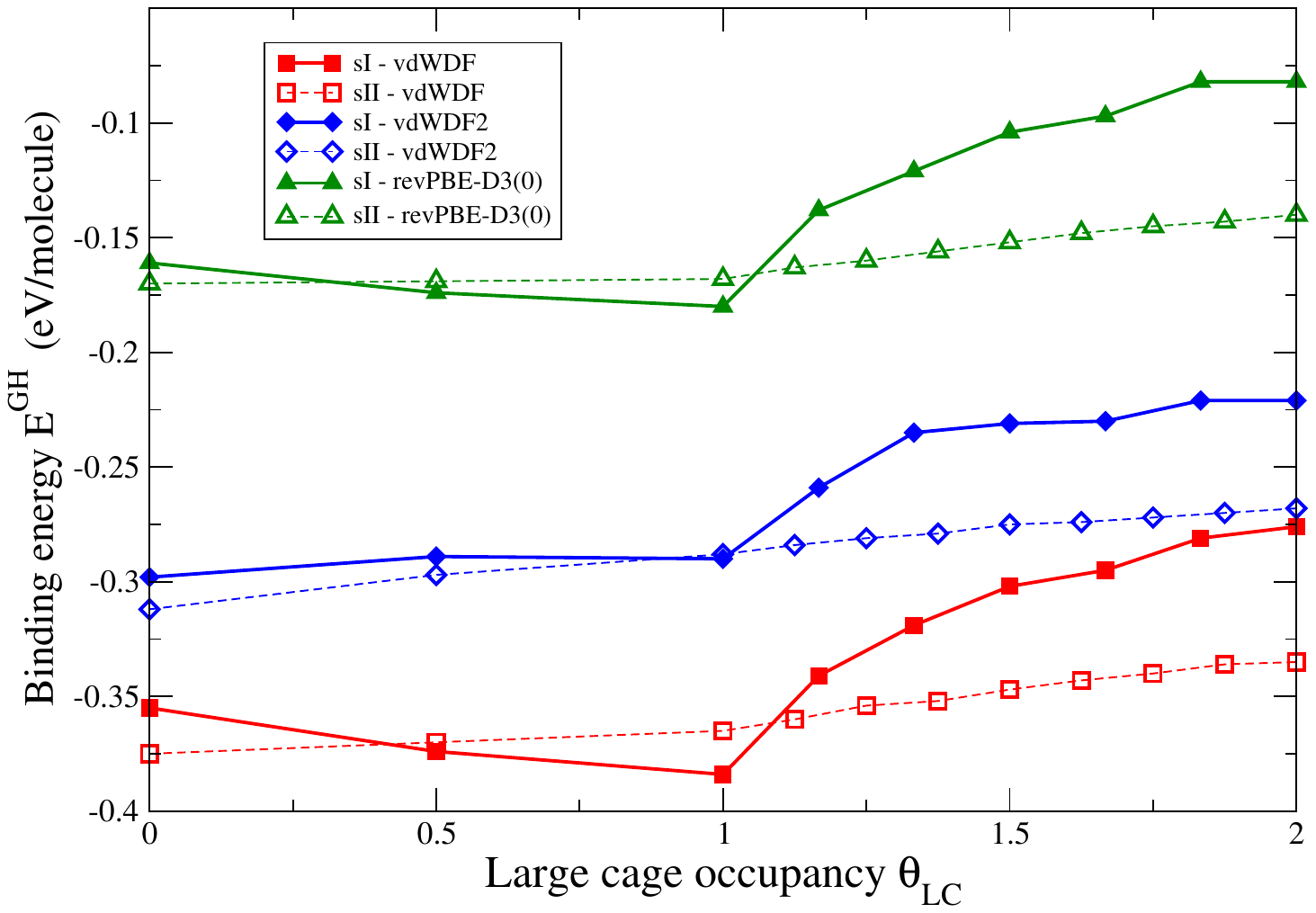}
        \label{fig:EGH}
    \end{subfigure}
    \hfill
    \begin{subfigure}[b]{0.49\textwidth}
        \centering
        \includegraphics[trim=15 25 75 90, clip, width=\textwidth]{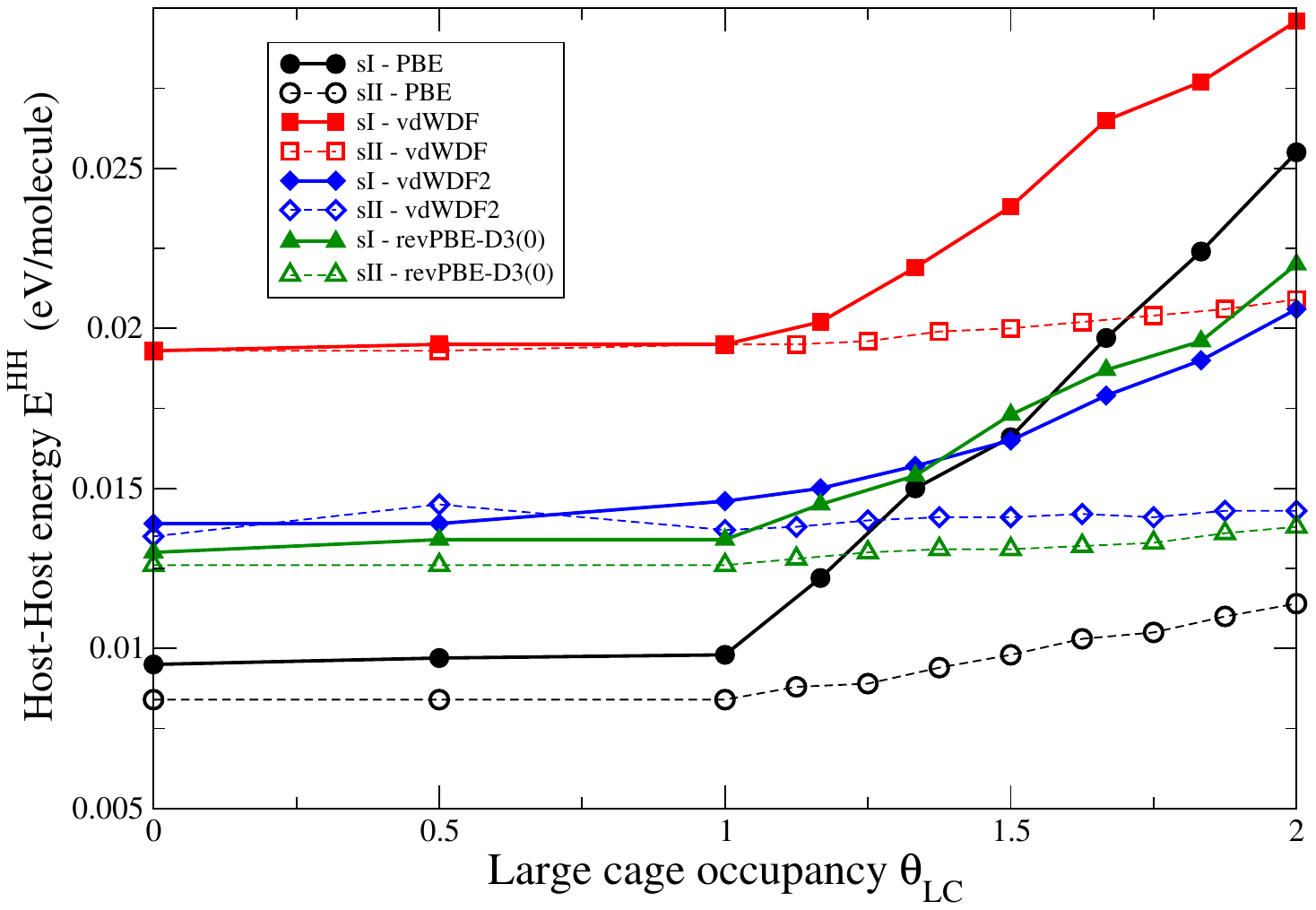}
        \label{fig:EHH}
    \end{subfigure}
    \caption{Guest–host energy $E^{GH}$ (left) and host–host energy $E^{HH}$ (right) as a function of the large cage occupancy $\theta_{LC}$, obtained with various DFT functionals: PBE (black circles), vdW-DF (red squares), vdW-DF2 (blue diamonds), and revPBE-D3(0) (green triangles) for sI (solid lines) and sII (dashed lines) structures.}
    \label{fig:Energies}
\end{figure}

Figure~\ref{fig:Energies} displays the evolution of the guest–host ($E^{\mathrm{GH}}$) and host–host ($E^{\mathrm{HH}}$) interaction energies as a function of the large-cage occupancy $\theta_{\mathrm{LC}}$ for the sI and sII N$_2$ clathrate structures. \rev{Results obtained with vdW-DF and vdW-DF2 are also reported for comparison, in order to assess the robustness of the energetic trends with respect to the choice of exchange-correlation functional, beyond the structural validation provided by revPBE-D3(0).} The PBE functional is not shown in Figure~\ref{fig:Energies} (left panel) since the corresponding guest–host interaction energy remains positive (or near zero) over the entire range of occupancies, which is a direct consequence of the absence of long-range dispersion interactions in this functional. For the three dispersion-inclusive functionals (revPBE-D3(0), vdW-DF, and vdW-DF2), $E^{\mathrm{GH}}$ exhibits a similar behavior: it remains nearly constant or slightly decreases in the under-occupancy region ($\theta_{\mathrm{LC}}\!\leqslant\!1$) and increases in the double-occupancy domain, particularly for the sI structure. This rise reflects growing repulsive interactions between guest molecules confined in smaller large cages. As already noted in Metais~\cite{Metais2021}, the binding energy becomes lower for sII compared with sI when the large cages are doubly occupied. This confirms, regardless of the functional employed, that double occupancy stabilizes the sII framework from the $E^{\mathrm{GH}}$ point of view, in agreement with previous experimental and theoretical studies~\cite{Zhu2014,Petuya2018a,Petuya2019,Metais2021}.

A comparison of absolute binding-energy values indicates that revPBE-D3(0) provides the best agreement with meta-GGA functionals of the M06 family, which themselves reproduce post-Hartree–Fock and experimental reference data for H$_2$O and H$_2$O/CH$_4$ interactions in clathrate hydrates~\cite{Liu2012,Sun2017}. For instance, the M06-2X functional predicts an average binding energy for single occupancy of isolated cages of sI/sII N$_2$ clathrate of about 0.22~eV per molecule~\cite{Sun2017}, in close agreement with revPBE-D3(0), which yields $\sim\SI{0.18}{\electronvolt}$ per molecule at $\theta_{\mathrm{LC}}=1$.

Two distinct regimes are also observed for the host–host contribution $E^{\mathrm{HH}}$: it remains almost unaffected by occupancy up to $\theta_{\mathrm{LC}}\!=\!1$, then increases once double occupancy occurs. This effect is more pronounced for sI, owing to its smaller large cages, whereas sII can accommodate two guest molecules with less distortion of the water network. Among the functionals, PBE induces the strongest framework perturbation because the missing dispersion attraction cannot compensate for the repulsive deformation energy. In contrast, vdW-DF2 and revPBE-D3(0) yield nearly identical trends and are expected to provide a more realistic description of water–water interactions in clathrate hydrates~\cite{Liu2012,Cox2014}.

\begin{figure}[ht]
  \centering
    \begin{subfigure}[b]{0.49\textwidth}
        \centering
        \includegraphics[trim=5 25 75 90, clip, width=\textwidth]{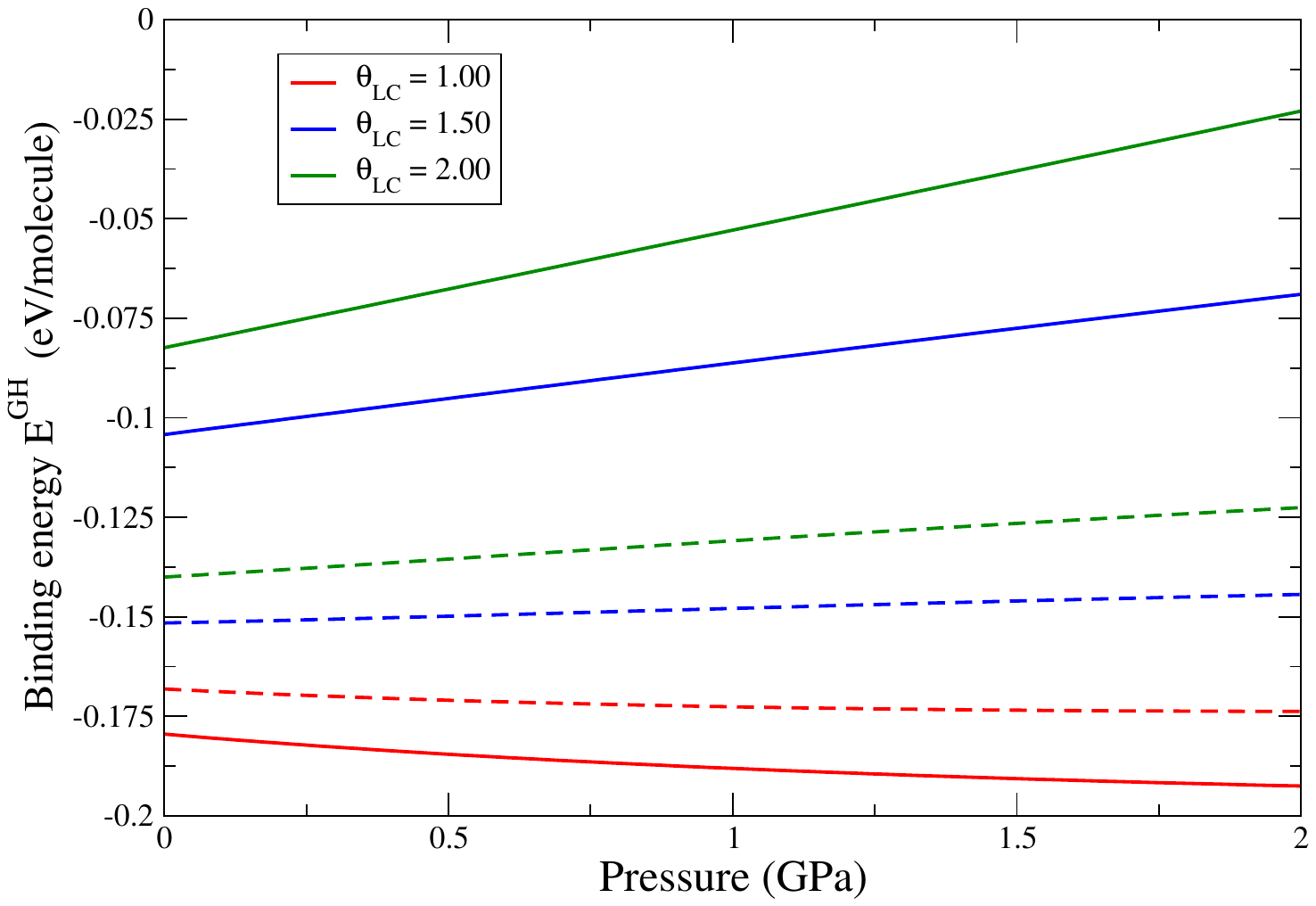}
        \label{fig:EGHPres}
    \end{subfigure}
    \hfill
    \begin{subfigure}[b]{0.49\textwidth}
        \centering
        \includegraphics[trim=5 25 75 90, clip, width=\textwidth]{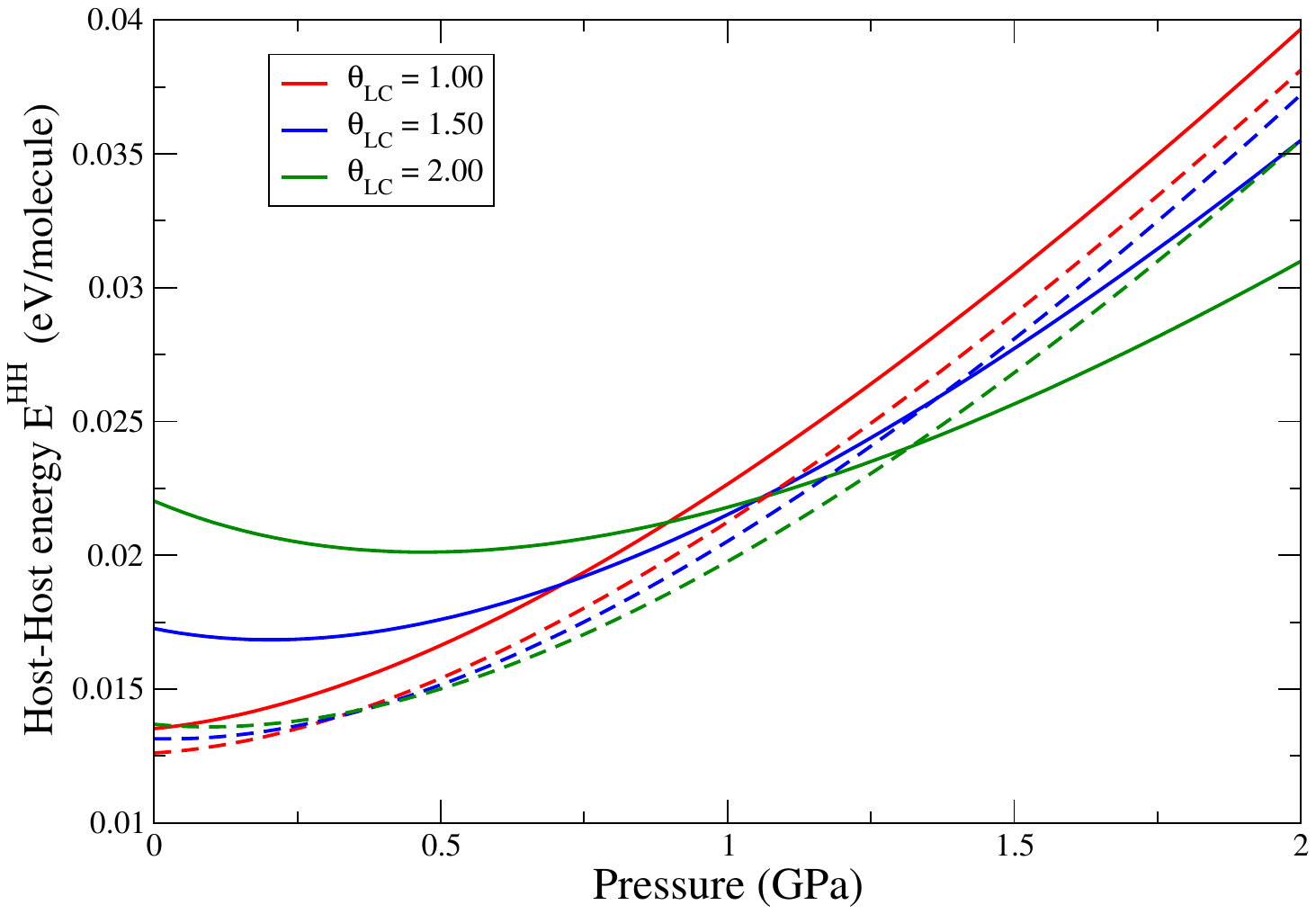}
        \label{fig:EHHPres}
    \end{subfigure}
    \caption{Guest–host energy $E^{GH}$ (left) and host–host energy $E^{HH}$ (right) as a function of pressure in GPa, obtained with the revPBE-D3(0) functional for different large cage occupancies $\theta_{LC}$. Solid and dashed lines correspond to sI and sII structures, respectively.}
    \label{fig:EnergiesPres}
\end{figure}

Evolution of $E^{\mathrm{GH}}$ and $E^{\mathrm{HH}}$ interaction energies with pressure is shown in Figure~\ref{fig:EnergiesPres} for three different large-cage occupancies $\theta_{\mathrm{LC}}$. As in Figure~\ref{fig:unitcellPres}, the analysis is based on revPBE-D3(0) calculations. Regarding the guest–host interaction energy, it becomes increasingly negative with pressure for single occupancy ($\theta_{\mathrm{LC}}=1$), reflecting the enhancement of guest–host interactions as the cages contract and the N$_2$–H$_2$O distances shorten. For $\theta_{\mathrm{LC}}>1$, an opposite trend is observed: $E^{\mathrm{GH}}$ increases with pressure, and the increase becomes steeper with higher occupancies. This behavior is more pronounced for sI, where the more confined large cages impose stronger spatial constraints on the guest molecules, whereas sII maintains a smoother variation of $E^{\mathrm{GH}}$ owing to its more open framework. For under-occupied configurations ($\theta_{\mathrm{LC}}<1$ or $\theta_{\mathrm{SC}}<1$), not shown here, $E^{\mathrm{GH}}$ decreases sharply when the small cages are emptied but increases when the large cages are fully vacant, confirming that stabilization mainly originates from N$_2$–H$_2$O interactions in the large cages (see Figures S1 and S2 in the Supporting Information).

The host–host interaction energy $E^{\mathrm{HH}}$ shows an opposite trend: it increases monotonically with pressure, representing the growing elastic cost associated with framework compression and hydrogen-bond deformation, although the rate of increase becomes progressively weaker with increasing occupancy. In particular, single occupancy, which corresponds to the lowest $E^{\mathrm{HH}}$ at zero pressure, yields the highest values at $P=\SI{2}{\giga\pascal}$. This evolution reflects the redistribution of stress between the guest and the host framework. Indeed, for singly occupied cages, the external pressure is mainly transmitted to the host lattice, which accommodates the stress through hydrogen-bond bending and O···O shortening, resulting in a larger increase of $E^{\mathrm{HH}}$. For doubly occupied cages, the internal pressure exerted by the two confined N$_2$ molecules counteracts the external compression, reducing host deformation and the corresponding increase of $E^{\mathrm{HH}}$. This internal cushioning effect of the guest molecules is particularly visible in the sI phase, where the smaller large cages are more sensitive to the presence of internal support.

\begin{figure}[ht]
  \centering
    \begin{subfigure}[b]{0.49\textwidth}
        \centering
        \includegraphics[trim=5 25 75 90, clip, width=\textwidth]{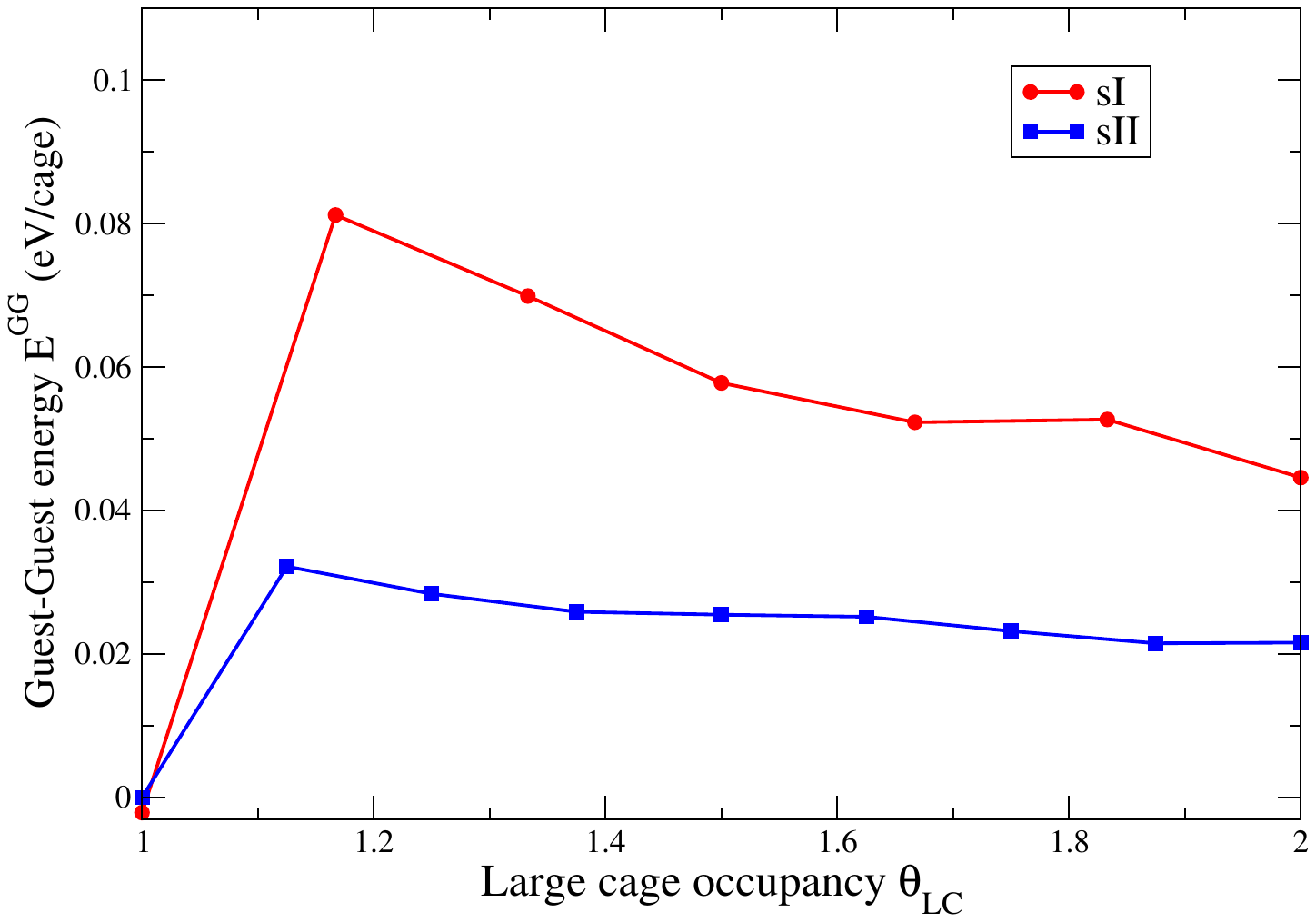}
        \label{fig:EGG}
    \end{subfigure}
    \hfill
    \begin{subfigure}[b]{0.49\textwidth}
        \centering
        \includegraphics[trim=5 25 75 90, clip, width=\textwidth]{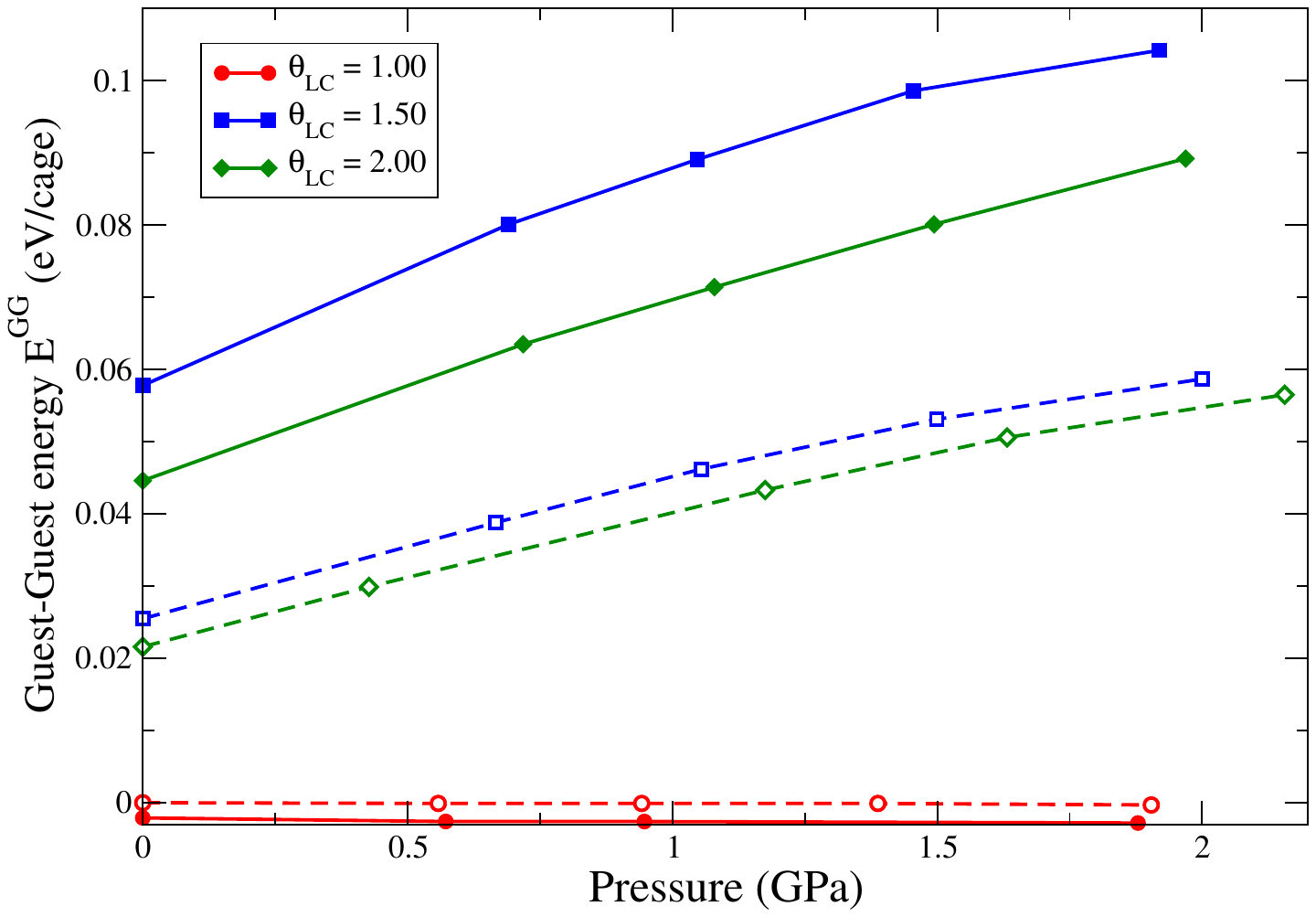}
        \label{fig:EGGPres}
    \end{subfigure}
    \caption{Guest–guest interaction energy $E^{GG}$ as a function of the large cage occupancy $\theta_{LC}$ (left) and as a function of pressure (right), obtained with the revPBE-D3(0) functional. In the right figure, solid and dashed lines correspond to sI and sII structures, respectively.}
    \label{fig:Eguestguest}
\end{figure}

\rev{The guest--guest interaction energy $E^{\mathrm{GG}}$ provides further insight into the role of confinement and occupancy (Figure~\ref{fig:Eguestguest}). For singly occupied structures ($\theta_{\mathrm{LC}}=1$), $E^{\mathrm{GG}}$ remains negligible, indicating that intermolecular interactions between guests located in different cages are very weak. In contrast, for doubly occupied large cages, $E^{\mathrm{GG}}$ becomes positive, reflecting a repulsive interaction between the two confined \chemform{N_2} molecules. This repulsion is significantly stronger in the sI structure than in sII, consistent with the smaller size of the sI large cages and the resulting stronger confinement. Interestingly, $E^{\mathrm{GG}}$ decreases as soon as more than one large cage is doubly occupied, indicating that the host framework progressively adapts to accommodate two guest molecules within the same cage. This reduction of the effective repulsion can be attributed to a progressive distortion of the water network, which increases the available free volume and allows a more favorable spatial arrangement of the guests. This interpretation is consistent with the simultaneous increase of $E^{\mathrm{HH}}$ and the lattice expansion observed in Figures~\ref{fig:unitcell} and~\ref{fig:Energies}.}

\rev{The pressure dependence of $E^{\mathrm{GG}}$ (Figure~\ref{fig:Eguestguest}, right) further highlights the role of confinement: $E^{\mathrm{GG}}$ increases monotonically with pressure for doubly occupied cages, as expected from the enhanced steric repulsion induced by cage compression. This trend is again more pronounced for sI than for sII, confirming that the smaller cages are more sensitive to external compression. These results clearly demonstrate that guest--guest interactions are negligible in the single-occupancy regime but become a key repulsive contribution in doubly occupied cages, strongly modulated by both cage size and host flexibility.}

Overall, the combined evolution of $E^{\mathrm{GH}}$ and $E^{\mathrm{HH}}$ with pressure reveals that the stabilization of N$_2$ clathrate hydrates results from a delicate balance between guest–host attraction and host-framework elasticity, both strongly dependent on structure and cage filling. The next section therefore examines how these competing contributions translate into overall stability by constructing pressure-dependent convex-hull diagrams.

\subsection{Phase stability with pressure}
\begin{figure}[ht]
  \begin{center}
    \includegraphics[width=0.98\textwidth]{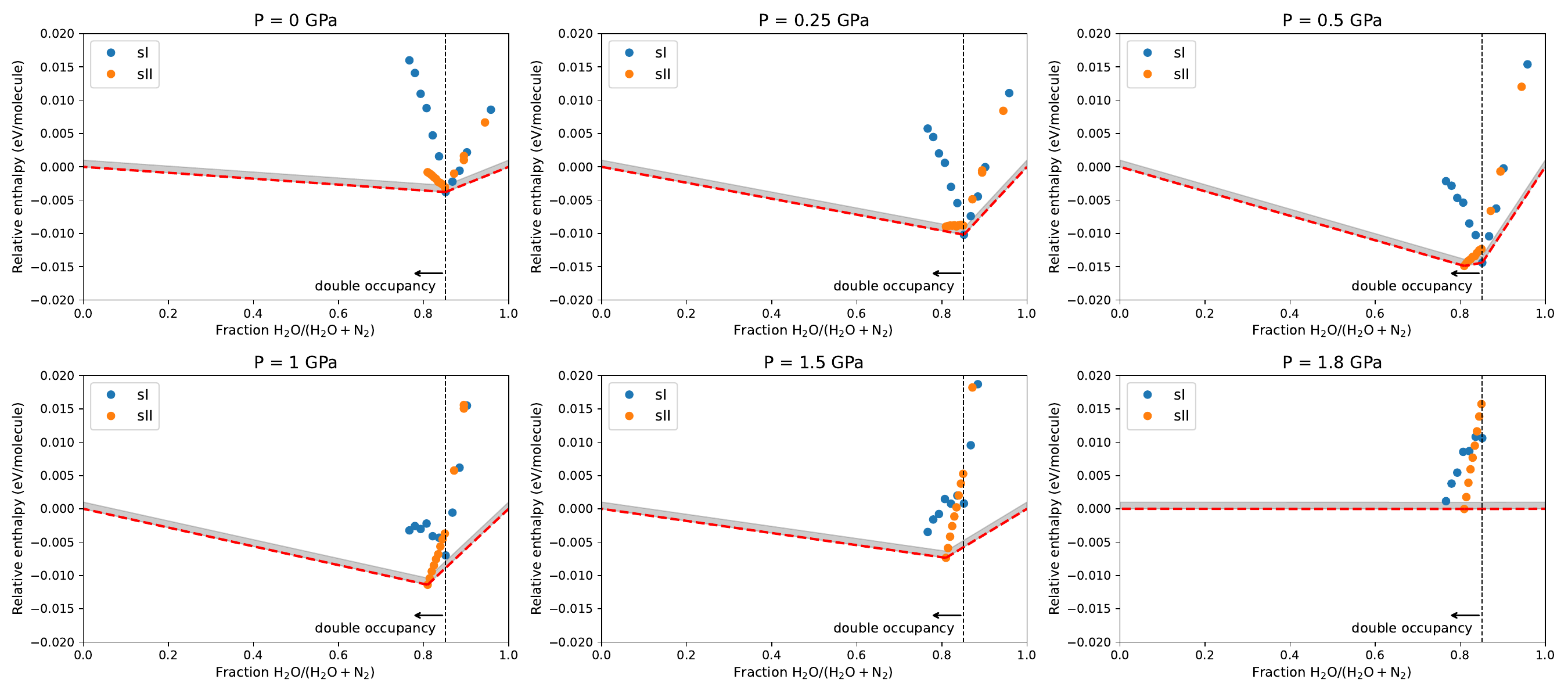}
    \caption{Convex-hull diagram for \chemform{H_2O\text{--}N_2} clathrate hydrate obtained with the revPBE-D3(0) functional at selected pressures and zero temperature. Relative enthalpy of formation (in eV/molecule) is plotted as a function of the \chemform{H_2O} fraction. Reference compounds are pure \chemform{H_2O} ice and \chemform{N_2}, calculated at 0K for the indicated pressure. Blue and orange circles correspond to sI and sII structures, respectively. The vertical dashed lines indicate the position of single occupancy of cages; all points to the left of this line correspond to double occupancy. Shaded grey regions above the convex hull ($\SI{\sim 1}{\milli\electronvolt}$/molecule) take into account the DFT numerical uncertainties.}
    \label{fig:convexhull}
  \end{center}
\end{figure}

The following section focuses on the revPBE-D3(0) results, in line with the previous discussion; results obtained with other functionals are reported in the Supporting Information for comparison. The total cohesive behavior arises from the balance between the guest–host ($E^{\mathrm{GH}}$) and host–host ($E^{\mathrm{HH}}$) interaction energies through the evaluation of the non-bonding energy $E^{\mathrm{NB}}$, calculated from Eq.~\ref{eq:NBsI}–\ref{eq:NBsII} at zero pressure and Eq.~\ref{eq:DeltaH}–\ref{eq:DeltaHcompact} when pressure is included. Convex-hull plots are shown in Figure~\ref{fig:convexhull} for N$_2$ clathrate hydrates at $T=\SI{0}{\kelvin}$, where the relative enthalpy with respect to pure ice and solid nitrogen is plotted as a function of the water fraction for several pressures. Results for pure ice and solid nitrogen are provided in the Supporting Information for the revPBE-D3(0) functional. The stability sequence of hydrogen-ordered ice phases is consistent with experimental data~\cite{Salzmann2019}, confirming that revPBE-D3(0) provides a reliable description of hydrogen-bonded systems. However, for solid nitrogen at $T=\SI{0}{\kelvin}$, where the experimentally stable phases are $\alpha$, $\gamma$, and $\varepsilon$~\cite{Tassini2005}, the DFT predictions do not fully reproduce the correct stability ordering. Although $\alpha$ and $\gamma$ are indeed stable at low pressure, the $\varepsilon$ phase does not become favorable near $\SI{2}{\giga\pascal}$ as observed experimentally. This limitation of DFT, previously reported for other functionals~\cite{Pickard2009}, was corrected by post-Hartree–Fock calculations~\cite{Erba2011}. In our study, the $\gamma$ phase remains the most stable up to $\SI{\sim 2}{\giga\pascal}$ and is therefore used as the solid-nitrogen reference.

\begin{figure}[ht]
  \begin{center}
    \includegraphics[width=0.9\textwidth]{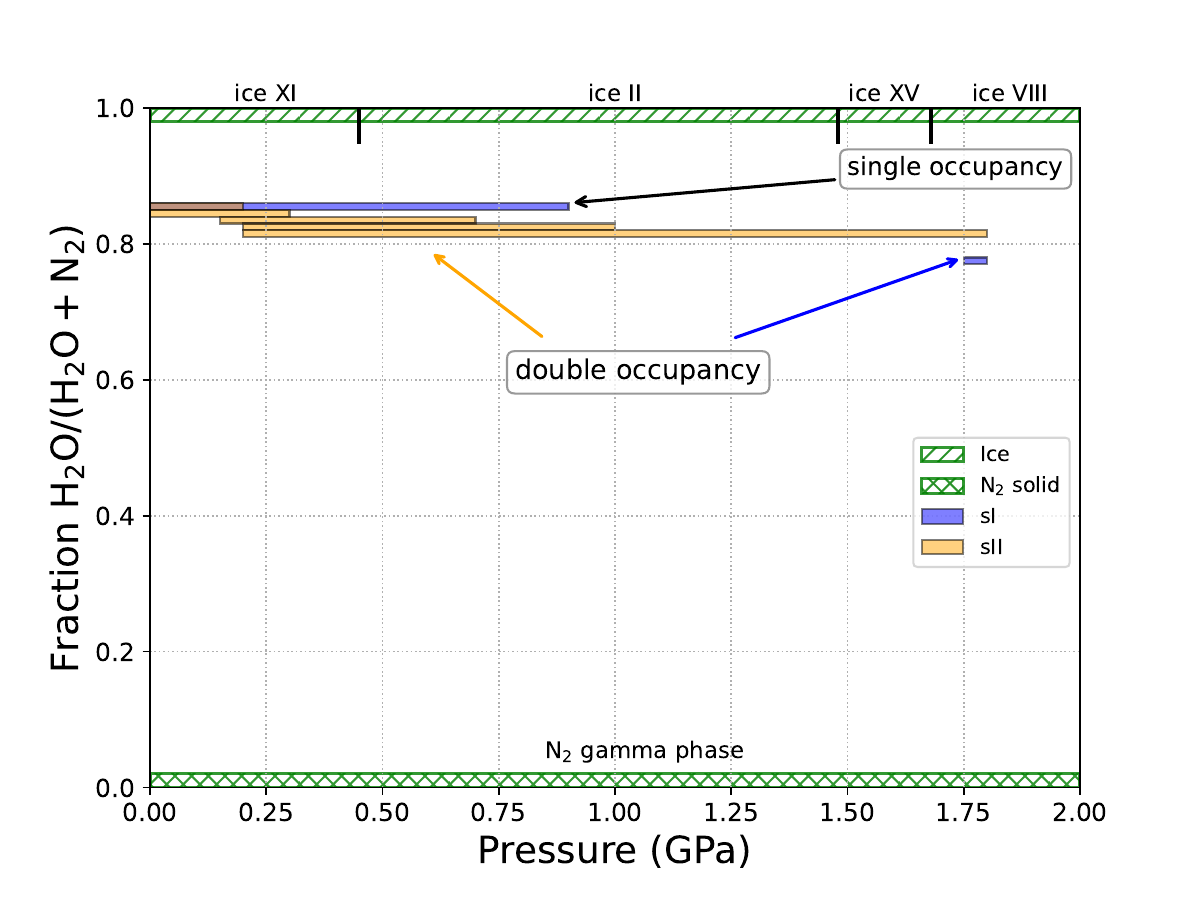}
    \caption{Phase diagram of the \chemform{H_2O\text{--}N_2} clathrate hydrate obtained with the revPBE-D3(0) functional at zero temperature. The stability areas of sI (blue) and sII (orange) clathrate hydrate structures are shown as a function of pressure and \chemform{H_2O} fraction. Stable ice phases obtained with the functional are indicated on top, and the \chemform{N_2} solid phase is indicated on the bottom. The upper rectangle corresponds to single occupancy and all other rectangles represent double occupancy.}
    \label{fig:diagphase}
  \end{center}
\end{figure}

At low pressure (\SI{0}{\giga\pascal}), both sI and sII clathrates with single occupancy (SO) lie on the convex hull, indicating thermodynamic stability relative to separated ice and solid nitrogen. Upon compression to 0.25~GPa, the sI single-occupancy point remains on the hull, while sII compositions with progressively filled large cages (from $\theta_{\mathrm{LC}}=1$ to 1.8) become nearly degenerate within the $\SI{1}{\milli\electronvolt}$ uncertainty range, suggesting the onset of stabilization of double occupancy (DO). At around \SI{1}{\giga\pascal}, the double-occupancy sII compositions clearly fall on the convex hull, whereas the single-occupancy sI points shift upward, marking a pressure-induced preference for the denser sII framework with multiple occupancy. At the highest investigated pressure ($\SI{\sim1.8}{\giga\pascal}$), the double-occupancy sI and sII points converge to similar relative enthalpies, both approaching $\Delta H \approx 0$, indicating increasing competition between clathrate stability and dissociation into the separate solid phases. Thus, while single occupancy is favored for both structures at ambient pressure, increasing pressure progressively stabilizes sII double occupancy, whereas sI with single occupancy remains stable up to about \SI{0.8}{\giga\pascal}.These pressure-dependent trends are summarized in the compositional phase diagram of Figure~\ref{fig:diagphase}, which delineates the stability domains of sI and sII as a function of cage occupancy and pressure.

Despite sI often being considered experimentally metastable at low–moderate pressures, the present results show that the single-occupancy sI structure remains thermodynamically stable up to about $\SI{0.8}{\giga\pascal}$. In contrast, sII displays a broader stability range: single occupancy is stabilized first as pressure rises, followed by double occupancy at higher pressures, reflecting the larger large-cage volume and reduced framework strain in sII. At still higher pressures ($\gtrsim\SI{1.75}{\giga\pascal}$), the diagram reveals a narrow window where sI reappears, in close agreement with the van der Waals–Platteeuw model prediction of sI stability at $T=0$~K above $\SI{1}{\giga\pascal}$~\cite{Lundgaard1992}. However, this stability is marginal relative to separation into the pure solid phases (ice and $\mathrm{N_2}$), as indicated by its proximity to the convex-hull tie line connecting the pure components.

A similar global topology of the phase map is obtained with nonlocal vdW-DF-type functionals (see Supporting Information), although the SO$\to$DO and sI$\leftrightarrow$sII transition lines occur at somewhat lower pressures. This shift reflects the smaller $B_0'$ values predicted by these functionals, i.e. a weaker stiffening of the framework with pressure, which favors earlier stabilization of double occupancy. We emphasize that the present diagram is a $T=0$~K thermodynamic limit. At finite temperature, entropic contributions associated with guest distribution among cages, molecular packing, and lattice vibrations are expected to influence the overall stability of N$_2$ clathrate hydrates~\cite{Sun2017}. Finally, the current map does not include other possible host frameworks such as hexagonal clathrate or filled-ice phases; these additional topologies can compete at elevated pressures and would further refine the high-pressure part of the diagram if considered explicitly.

\section{Conclusions}
A pressure–composition thermodynamic framework has been established for \chemform{N_2} clathrate hydrates from first principles. Systematic DFT calculations, combined with Birch–Murnaghan equation-of-state fitting and convex-hull analyses, were employed to quantify the influence of lattice occupancy and pressure on the stability competition between the sI and sII structures. Among the tested functionals, revPBE-D3(0) was found to provide the most accurate description, reproducing experimental lattice parameters and bulk moduli $B_0$ while yielding realistic pressure derivatives $B_0'$. The guest–host and host–host interaction analyses demonstrate the strong energetic impact of large-cage double occupancy, especially in sI, and reveal the redistribution of mechanical stress within the water framework under compression.

Convex-hull constructions indicate that both sI and sII with single occupancy remain stable up to approximately $\SI{0.8}{\giga\pascal}$, whereas increasing pressure progressively stabilizes the double-occupancy sII configuration, consistent with its larger large-cage volume and reduced framework strain. At higher pressures, sI and sII double-occupancy configurations approach similar enthalpies, competing with phase separation into pure ice and solid \chemform{N_2}.

The present $T=0$~K framework does not account for entropic contributions, which are expected to influence hydrate stability at finite temperature, nor does it include alternative host frameworks (such as hexagonal clathrates or filled-ice phases) that may compete under compression. Future extensions of this framework to finite temperature, using quasi-harmonic or anharmonic free-energy calculations, and to other candidate structures will allow the construction of quantitatively predictive stability diagrams for \chemform{N_2} and mixed-guest hydrates under realistic thermodynamic conditions.

\section*{Supporting Information}
Additional structural figures including Birch--Murnaghan fitting parameters for various exchange-correlation functionals, intermolecular energy analysis, relative enthalpy of pure \chemform{N_2} and ice, convex-hull and phase diagrams (PDF).\\
Optimized atomic coordinates of representative \chemform{N_2} clathrate hydrate structures provided as VASP POSCAR files (ZIP).

\section*{Acknowledgments}
This paper falls in the frame of the PhySH project ANR-22-CE50-0003-01 funded by the French ANR (Agence Nationale de la recherche). A.P. acknowledges support from PID2022-140845OB-C66 funded by MCIN/AEI/10.13039/501100011033 and FEDER, EU.

This work was provided with computing HPC and storage resources by the Mésocentre de Calcul at the Université de Franche-Comté and by GENCI at TGCC thanks to the grant 2025-A0180816220 on the supercomputer Joliot Curie's ROME partition.

\bibliography{ClathN2_CH}
\clearpage

\section*{TOC Graphic}
\begin{figure}[h!]
 \centering
 \includegraphics[width=8.25cm]{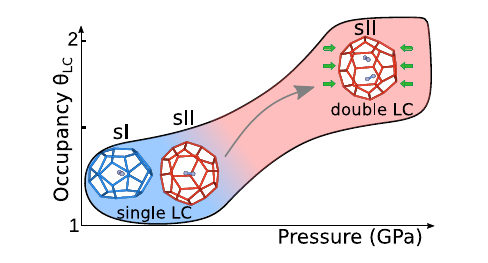}
\end{figure}

\end{document}